\documentclass[12pt]{article}
\usepackage[latin1]{inputenc}
\usepackage{blkarray}
\usepackage{lipsum}
\usepackage{amsmath,amssymb,amsthm}
\usepackage{physics}
\usepackage{color}
\usepackage{amsfonts}
\usepackage[mathscr]{euscript}
\usepackage{graphicx}
\usepackage{geometry}
\usepackage{amssymb,epsfig}
\usepackage{subcaption}
\usepackage{comment}
\usepackage{tabularx}
\usepackage{bm}
\usepackage{euscript}
\usepackage[dvipsnames]{xcolor}
\usepackage{amsfonts}
\usepackage{exscale}
\usepackage{amsbsy}
\usepackage{textcomp}
\usepackage{hyperref}
\usepackage{slashed}
\usepackage{authblk}
\usepackage{tabularx}
\usepackage{euscript}
\usepackage{graphicx}
\usepackage{color}
\usepackage{exscale}
\usepackage{amsbsy}
\usepackage{textcomp}
\usepackage{hyperref}
\usepackage{doi}
\usepackage{tensor}
\usepackage{wrapfig}
\usepackage[font=footnotesize,labelfont=bf]{caption}
\usepackage{ulem} 
\usepackage{makecell}
\usepackage[mathscr]{euscript}

\def\ba#1\ea{\begin{align}#1\end{align}}
\def\bg#1\eg{\begin{gather}#1\end{gather}}
\def\bm#1\em{\begin{multline}#1\end{multline}}
\def\bmd#1\emd{\begin{multlined}#1\end{multlined}}

\newcommand{\be}{\begin{equation}}
	\newcommand{\ee}{\end{equation}}
\newcommand{\bea}{\begin{eqnarray}}
	\newcommand{\eea}{\end{eqnarray}}

\newcommand{\bs}{\boldsymbol}
\newcommand{\matleft}{\left(\begin{array}}
	\newcommand{\matright}{\end{array}\right)}
\newcommand{\sgn}{\operatorname{sgn}}

\usepackage[numbers,sort&compress]{natbib}
\def\simge{
	\mathrel{\rlap{\raise 0.511ex 
			\hbox{$>$}}{\lower 0.511ex \hbox{$\sim$}}}}

\def\simle{
	\mathrel{\rlap{\raise 0.511ex 
			\hbox{$<$}}{\lower 0.511ex \hbox{$\sim$}}}}

\makeatletter
\renewcommand\section{\@startsection {section}{1}{\z@}%
	{-3.5ex \@plus -1ex \@minus -.2ex}
	{2.3ex \@plus.2ex}%
	{\normalfont\large\bfseries}}
\renewcommand\subsection{\@startsection{subsection}{2}{\z@}%
	{-3.25ex\@plus -1ex \@minus -.2ex}%
	{1.5ex \@plus .2ex}%
	{\normalfont\bfseries}}
\renewcommand\subsubsection{\@startsection{subsubsection}{3}{\z@}%
	{-3.25ex\@plus -1ex \@minus -.2ex}%
	{1.5ex \@plus .2ex}%
	{\normalfont\itshape}}
\makeatother

\def\pplogo{\vbox{\kern-\headheight\kern -29pt
		\halign{##&##\hfil\cr&{\ppnumber}\cr\rule{0pt}{2.5ex}&\ppdate\cr}}}
\makeatletter
\def\ps@firstpage{\ps@empty \def\@oddhead{\hss\pplogo}%
	\let\@evenhead\@oddhead 
}
\hypersetup{
	unicode=false,          
	pdftoolbar=true,        
	pdfmenubar=true,        
	pdffitwindow=false,     
	pdfstartview={FitH},    
	pdftitle={Controlled_NFLtransport},    
	pdfauthor={},     
	pdfsubject={Subject},   
	pdfcreator={},   
	pdfproducer={}, 
	pdfkeywords={keyword1} {key2} {key3}, 
	pdfnewwindow=true,      
	colorlinks=true,       
	linkcolor=OliveGreen, 
	citecolor=NavyBlue,        
	filecolor=magneta,      
	urlcolor=cyan           
}

\numberwithin{equation}{section}

\newcommand*\samethanks[1][\value{footnote}]{\footnotemark}

\newcommand\beal{\begin{equation}\begin{aligned}}
		\newcommand\eeal{\end{aligned}\end{equation}}

\begin{document}

\normalem

\setcounter{page}0
\def\ppnumber{\vbox{\baselineskip14pt
}}

\def\ppdate{
} 
\date{}

\title{\Large\bf Controlled expansion for transport \\ in a class of non-Fermi liquids}
\author{Zhengyan Darius Shi}
\affil{\it\small Department of Physics, Massachusetts Institute of Technology, Cambridge, MA 02139, USA}
\maketitle\thispagestyle{firstpage}

\begin{abstract}

Non-Fermi liquids arise when strong interactions destroy stable fermionic quasiparticles. The simplest models featuring this phenomenon involve a Fermi surface coupled to fluctuating gapless bosonic order parameter fields, broadly referred to as ``Hertz-Millis" models. We revisit a controlled approach to Hertz-Millis models that combines an expansion in the inverse number ($N$) of fermion species with an expansion in the deviation of the boson dynamical critical exponent $z$ from 2. The structure of the expansion is found to be qualitatively different in the quantum critical regime $\Omega \ll q$ and in the transport regime $\Omega \gg q$. In particular, correlation functions in the transport regime involve infinitely many diagrams at each order in perturbation theory. We provide an explicit and tractable recipe to classify and resum these diagrams. For the simplest Hertz-Millis models, we show that this recipe is consistent with non-perturbative anomaly arguments and correctly captures the fixed point optical conductivity as well as leading corrections from irrelevant operators. We comment on potential applications of this expansion to transport in more complicated Hertz-Millis models as well as certain beyond-Landau metallic quantum critical points. 

\end{abstract}

\pagebreak
{
\hypersetup{linkcolor=black}
\tableofcontents
}
\pagebreak

\section{Introduction}

Over the past several decades, a growing number of metals have shown striking deviations from the predictions of Landau Fermi liquid theory, one of the cornerstones of modern condensed matter physics. Examples include the normal state of cuprate high temperature superconductors~\cite{Gurvitch1987_YBCO,Martin1990_Bi2201,VanderMarel_conductivityscaling_2003,Cooper2009_LSCO,Collignon2017_NdLSCO,Legros2019,Michon2022}, heavy fermion metals near a quantum critical point~\cite{Seaman1991_heavyNFL,Lohneysen1994_heavyNFL,Trovarelli2000_heavyNFL,Prochaska_conductivityscaling_2020,shen2020_heavyNFL}, and more recently twisted bilayer graphene~\cite{Cao2020,jaoui2022_TBG_Tlinear,Lyu2021_TBG_Hallangle}. The most robust feature shared by these materials is a $T$-linear resistivity down to the lowest accessible temperatures, in contrast to the $T^2$ resistivity seen in conventional Fermi liquids. Such ``strange metallic" transport signals the absence of stable electronic quasiparticles on the Fermi surface. Given that different strange metals have distinct proximate phases and phase transitions, there is likely not a single unifying effective theory. However, the destruction of quasiparticles requires strong interactions between the fermions, which must be an essential ingredient in any low energy description. 

A simple and natural route towards strong interactions is to couple the Fermi surface to gapless bosonic modes. When the bosons are mutually independent and only interact with the Fermi surface, these models describe Landau symmetry-breaking transitions in metals with the bosons playing the role of coarse-grained order parameter fields (often referred to as Hertz-Millis models after the pioneering works of~\cite{Hertz1976,Millis1993}). When there are multiple bosons with mutual interactions, we get a wider class of models that can encompass certain beyond-Landau metallic quantum critical points (e.g. the Kondo breakdown critical point proposed for heavy fermion metals~\cite{Si2001_Kondolattice,Si2003_Kondolattice,Zhu2003_Kondolattice,SunKotliar2003_Kondolattice,senthil2003_FLstar,senthil2004_Kondo}). In this paper, we will derive most of our results in the simpler context of Hertz-Millis models with scalar bosons and comment on generalizations towards the end. 

In 2+1 dimensions, Hertz-Millis models generically flow to strong coupling in the infrared limit, thereby invalidating a direct perturbative expansion in powers of the coupling constant. This fundamental difficulty prompted the development of various deformations of the Hertz-Millis model that introduce an additional small parameter, such that the original Hertz-Millis action is recovered when the parameter is order one~\cite{Lee1992,Kim1994,Nayak1994,Lee2009,Mross2010,Metlitski2010,Metlitski2010a,DalidovichLee2013,Fitzpatrick2013,Raghu2015,AguileraDamia2019,Esterlis2019,Esterlis2021,Ye2021,guo2022_largeN,Patel2022}. Although many critical singularities can be calculated to leading orders in these expansions, other important physical properties like the electrical conductivity are more difficult to compute. 

\begin{figure}[ht]
    \centering
    \includegraphics[width = 0.6 \linewidth]{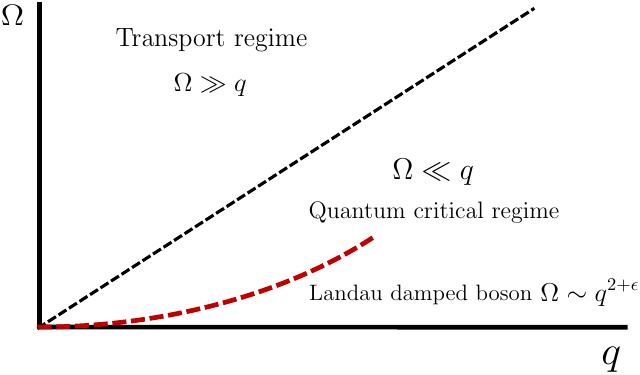}
    \caption{In a Hertz-Millis model, correlators in the quantum critical regime $(\Omega \ll q)$ control thermodynamic susceptibilities and critical exponents, while correlators in the transport regime $(\Omega \gg q)$ determine the conductivities of conserved currents. The structure of any perturbative expansion can be qualitatively different in these two kinematic regimes, which is a central challenge we must confront in this paper.}
    \label{fig:crit_vs_trans}
\end{figure}

The main objective of this work is to provide an efficient framework for transport calculations within a controlled expansion that respects all the non-perturbative constraints from symmetries and anomalies. The essential challenge we have to confront is the failure of the low frequency and low momentum limit to commute. For a general bosonic correlation function in the Hertz-Millis model evaluated at external frequency $\Omega$ and momentum $q$, one can distinguish between the ``quantum critical regime" where $\Omega \ll q$ and the transport regime where $\Omega \gg q$. The former controls static susceptibilities and critical exponents, while the latter determines the uniform conductivity of various conserved currents (see Fig.~\ref{fig:crit_vs_trans}). Given an expansion in powers of some small parameter $1/N$, the set of Feynman diagrams that contribute to the correlation function at a fixed order in $1/N$ can be very different in the two regimes and an expansion which is controlled in one regime need not be controlled in the other. A concrete example of this phenomenon occurs for the boson self energy $\Pi(\bs{q}, \Omega)$ in the Yukawa-SYK expansion~\cite{Esterlis2019,aldape2020solvable,Yuxuan2020_YukawaSYK,Esterlis2021,guo2022_largeN,Patel2022,shi2023_loop}. In the quantum critical regime, the leading order diagram reproduces Landau damping effects which are weakly dressed by higher order corrections. But in the transport regime, the same diagram gives a vanishing $\Pi(\bs{q}=0, \Omega)$ as $\Omega \rightarrow 0$ which violates the non-perturbative anomaly constraints in~\cite{shi2023_loop} when the order parameter is odd under inversion. This example shows that the interplay between these two regimes is subtle in general and deserves a more careful study.


Motivated by the above phenomenon, we ask whether there exists a controlled expansion in which bosonic correlation functions can be reliably computed in both the ``quantum critical regime" and the ``transport regime". We answer this question in the affirmative for the ``double expansion" introduced by Mross et al. in~\cite{Mross2010}. This expansion features $N$ species of fermions that interact strongly with a critical boson whose kinetic term has a non-analytic momentum dependence $|\bs{q}|^{1+\epsilon}$. A solvable limit can be obtained by taking $N$ large, $\epsilon$ small, with the product $\epsilon N$ fixed. One can then construct a systematic expansion order by order in $\epsilon$ (or equivalently in $1/N$), where all calculations can be performed directly for the full Fermi surface without passing to a multi-patch approximation. Physically, the value of $\epsilon$ controls the spatial falloff of long-range fermion interactions mediated by the boson. The $\epsilon = 0$ limit corresponds to a $1/r$ Coulomb interaction that naturally appears in the Halperin-Lee-Read theory for composite Fermi liquids in quantum Hall systems~\cite{Halperin1992}. The $\epsilon = 1$ limit restores the analyticity of the boson kinetic term and identifies the boson with a standard Landau order parameter. 

In previous works~\cite{Mross2010,Metlitski2014}, it has been understood that this perturbative scheme provides an efficient and controlled expansion for physical properties in the ``quantum critical regime", including dynamical critical exponents, fermion anomalous dimensions, and $2k_F$ singularities.\footnote{Up to a potential double-log divergence at four-loop order identified in~\cite{Ye2021} that we will discuss in Section~\ref{subsec:infinitude}.} However, it was observed in~\cite{shi2022_gifts} that even to leading order in the $1/N$ expansion, the boson self energy $\Pi(\bs{q}, \Omega)$ in the ``transport regime" $\Omega \gg q$ contains an infinite number of diagrams, suggesting that there is no tractable expansion. In this work, we show that this is not the case by rewriting the boson self energy as a geometric series 
\begin{equation}
    \Pi(\bs{q}, \Omega) = \bra{g_{\bs{q}, \Omega}} W_G (1-K)^{-1} \ket{g_{\bs{q}, \Omega}} \,,
\end{equation}
where $\ket{g_{\bs{q}, \Omega}}$ denotes the Yukawa interaction vertex, and $W_G, K$ are kernels acting on pairs of particle-hole lines that will be precisely defined in Section~\ref{sec:transport_expansion}.
Surprisingly, we find that the infinite set of diagrams for $\Pi(\bs{q} =0, \Omega)$ get repackaged into finitely many diagrams for $K$ at each order in $1/N$. Moreover, we provide in Section~\ref{subsec:surgery_suture} a ``surgery + suture" recipe for constructing transport diagrams from quantum critical diagrams, thereby facilitating efficient calculations of the electrical conductivity. 

This formalism can be directly applied to several problems in metallic quantum criticality. In the simplest case of a Fermi surface coupled to a single order parameter field (e.g. nematic or loop current order), we show that the leading order diagrams in the transport regime reproduce the fixed point optical conductivity derived using non-perturbative anomaly arguments in~\cite{shi2022_gifts,shi2023_loop,Else2023_collisionless}. Systematic corrections due to irrelevant operators are captured by the same set of diagrams that appear to leading order in the Yukawa-SYK expansion, except that the dynamical critical exponent is set to $z = 2 + \epsilon$~\cite{Maslov_yudson_Chubukov_2011,Maslov2016,guo2023migdal,guo2023fluctuation,Maslov2023_NFLoptical}. As emphasized in~\cite{guo2023migdal}, the Yukawa-SYK expansion suffers from a potential instability for $z > z_c = 8/3$. Within the double expansion, this instability is avoided for perturbatively small $\epsilon$. The fate of this instability as $\epsilon$ is extrapolated back to 1 is reserved for future studies.

In more complicated Hertz-Millis models where the fermions carry spin (or more general internal degrees of freedom) and the boson couples to some of the spin components, we work out modifications to the structure of $K$ and identify the leading order diagrams that contribute to the spin-resolved conductivities. A detailed evaluation of these diagrams in~\cite{shi2024_exciton} gives critical incoherent transport in the spin channel at the IR fixed point. It is our hope that the basic analysis in this work provides a solid foundation for perturbative studies of electrical transport in more general non-Fermi liquids.

The rest of the paper is organized as follows. In Section~\ref{sec:mross_review}, we set up the general Hertz-Millis action and review the basic structure of the double expansion. This review is followed by the identification of an infinite number of diagrams that contribute to the boson self energy at leading order in the transport regime. In Section~\ref{sec:transport_expansion}, we proceed to classify this infinite set of diagrams and give a recipe for efficient calculations. The main result is summarized in Section~\ref{subsubsec:recipe}. In Section~\ref{sec:application}, we apply this diagrammatic classification to several concrete models and compare with results in the literature. We conclude in Section~\ref{sec:discussion} with the merits/potential caveats of our approach and point out some future directions to explore. 

\section{Review of the controlled expansion}\label{sec:mross_review}

\subsection{Setup and basic properties in the quantum critical regime}\label{subsec:mross_basic}

Consider a Fermi surface formed by fermions $f$ coupled to the long wavelength and low frequency fluctuations of some bosonic order parameter $\phi$. The Euclidean action for this general Hertz-Millis model takes the form
\begin{equation}\label{eq:HMaction}
    \begin{aligned}
        S &= S_f + S_{\phi} + S_{\rm int} \,, \\
        S_f &= - \int_{\bs{k}, \omega} f^{\dagger}(\bs{k}, \omega) \left[i \omega - \epsilon_{\bs{k}}\right] f(\bs{k}, \omega) \,, \\
        S_\phi &= \frac{1}{2} \int_{\bs{q}, \Omega} \phi^*_a(\bs{q},\Omega) \left[\lambda \Omega^2 + K |\bs{q}|^2 + m^2\right] \phi_a(\bs{q}, \Omega) \,, \\
        S_{\rm int} &= \int_{\bs{k}, \bs{q}, \omega, \Omega} g^a(\bs{k}, \bs{q}) \phi_a(\bs{q}, \Omega) f^{\dagger}(\bs{k} + \bs{q}/2, \omega + \Omega/2) f(\bs{k} - \bs{q}/2, \omega - \Omega/2) \,, 
    \end{aligned}
\end{equation}
where $\epsilon_{\bs{k}}$ is some generic fermion dispersion and $m^2$ is the boson mass which vanishes at the critical point. The Yukawa interaction $S_{\rm int}$ couples the boson $\phi_a$ to a fermion bilinear with a form factor $g^a(\bs{k}, \bs{q})$. Important examples include the Ising-nematic critical point with $g(\bs{k}, \bs{q}) \sim \cos k_x - \cos k_y$ and fermions coupled to emergent gauge fields with $g(\bs{k}, \bs{q}) \sim \bs{v}_F(\bs{k}) \times \bs{\hat q}$.

In 2+1 dimensions, the Yukawa coupling is a relevant perturbation to the Gaussian fixed point. Therefore, the low energy properties of this model cannot be reliably extracted from perturbation theory and a controlled calculation requires deformations of the model with additional small parameters. The simplest way to introduce a small parameter is to consider the model with $N$ identical species of fermions transforming in the fundamental representation of $U(N)$ and perform an expansion in powers of $1/N$. Though tremendously successful in zero-density quantum field theories, this ``fundamental large $N$'' approach was shown to be uncontrolled in the presence of a Fermi surface in the pioneering work of Sung-Sik Lee in~\cite{Lee2009}. The crux of the issue can be summarized as follows: within the fundamental large $N$ expansion, the leading order fermion and boson Green's functions take the form
\begin{equation}\label{eq:doubleexp_leadingorder}
    G(\bs{k}, i\omega) \sim \frac{1}{i\omega + \frac{C}{N} \sgn(\omega) |\omega|^{2/3} - \epsilon_{\bs{k}}} \,, \quad D(\bs{q}, i\Omega) \sim \frac{1}{|\bs{q}|^2 + \gamma \frac{|\Omega|}{|\bs{q}|}} \,,
\end{equation}
where $C, \gamma$ are constants. In the infrared limit, the boson self energy is $N$-independent and takes the familiar Landau damping form. However, the $\mathcal{O}(1/N)$ term in the fermion self energy dominates over the bare energy for $\omega \ll 1/N^3$ and the fermion Green's function is enhanced by a power of $N$ as $\omega \rightarrow 0$. When factors of $G$ enter higher order Feynman diagrams as internal propagators, this enhancement invalidates the naive $N$-counting, rendering the expansion uncontrolled. 

To get around this problem, Mross et al. introduced a further deformation of the Hertz-Millis action \eqref{eq:HMaction} inspired by earlier works of Nayak and Wilczek~\cite{Nayak1994}. Here, the $f$-sector still contains $N$ identical species of fermions, but the $\phi$-sector has a non-analytic kinetic term $q^{1+\epsilon}$. The full Euclidean action therefore takes the form
\begin{equation}\label{eq:HMaction_doubleexp}
    \begin{aligned}
        S &= S_f + S_{\phi} + S_{\rm int} \,, \\
        S_f &= - \int_{\bs{k}, \omega} \sum_{i=1}^N f^{\dagger}_i(\bs{k}, \omega) \left[i \omega - \epsilon_{\bs{k}}\right] f_i(\bs{k}, \omega) \,, \\
        S_\phi &= \frac{1}{2} \int_{\bs{q}, \Omega} \phi^*_a(\bs{q},\Omega) \left[\lambda \Omega^2 + |\bs{q}|^{1+\epsilon} + m^2\right] \phi_a(\bs{q}, \Omega) \,, \\
        S_{\rm int} &= \frac{1}{\sqrt{N}} \int_{\bs{k}, \bs{q}, \omega, \Omega} g^a(\bs{k}, \bs{q}) \phi_a(\bs{q}, \Omega) \sum_{i=1}^N f^{\dagger}_i(\bs{k} + \bs{q}/2, \omega + \Omega/2) f_i(\bs{k} - \bs{q}/2, \omega - \Omega/2) \,.
    \end{aligned}
\end{equation}
The proposal of~\cite{Mross2010} is that a $1/N$ expansion becomes controlled if $N \rightarrow \infty$ and $\epsilon \rightarrow 0$ with the product $\epsilon N$ fixed. This can be simply understood by examining the fermion self energy for general $\epsilon \in [0,1]$,
\begin{equation}
    \Sigma(\bs{k}, i\omega) \sim \frac{1}{N \sin \frac{2\pi}{2 +\epsilon}} \sgn(\omega) |\omega|^{\frac{2}{2+\epsilon}} \,. 
\end{equation}
In the limit of $\epsilon = 1$, this self energy scales as $1/N$, leading to the infrared problems identified in the fundamental large $N$ expansion. But when $\epsilon \rightarrow 0$ and $N \rightarrow \infty$, the self energy scales as
\begin{equation}
    \Sigma(\bs{k}, i\omega) \sim \frac{1}{N \epsilon} \sgn(\omega) |\omega|^{\frac{2}{2+\epsilon}} \,.
\end{equation}
Since $\epsilon N$ is held fixed, the prefactor in $\Sigma(\bs{k}, i\omega)$ is now $\mathcal{O}(1)$ and the infrared problems in the fundamental large $N$ expansion are avoided. 

Based on the logic above, we expect that higher order corrections to the fermion and boson self-energies organize themselves into the following series
\begin{equation}
    \Sigma(\bs{k}, i\omega) = \sum_{n=0}^{\infty} N^{-n} g_c^{(n)}(\epsilon N, \bs{k}, \omega) \,,\quad \Pi(\bs{q}, i\Omega) = \sum_{n=0}^{\infty} N^{-n} g_b^{(n)}(\epsilon N, \bs{q}, \Omega) \,,
\end{equation}
where $g_c^{(n)}, g_b^{(n)}$ are continuous functions that have finite limits as $N \rightarrow \infty, \epsilon \rightarrow 0$ and $\epsilon N$ fixed. At each order in $1/N$, these functions can be found by enumerating all diagrams according to their $N$-scaling in the fundamental large $N$ expansion and evaluating them. Having avoided the infrared singularities in the fundamental large $N$ expansion, we expect the dependence of $g_c^{(n)}$ and $g_b^{(n)}$ on frequency and momenta to be no more singular than the leading order results in \eqref{eq:doubleexp_leadingorder} (up to logarithms). This structure has been verified up to three loops in~\cite{Mross2010}. A recent tour de force calculation~\cite{Ye2021} shows that two four-loop diagrams involving virtual Cooper pairs contain a double-log divergence which may destabilize the expansion. Concretely, working within a pair of anti-podal patches on the Fermi surface with coordinates $x, y$ perpendicular/parallel to the patches, the double-log divergence is a contribution to the boson self energy that scales as $\Pi(q_x, q_y, i\Omega) \sim |q_y| (\log \Lambda_y/|q_y|)^2$. The authors of~\cite{Ye2021} argued that no other diagram can cancel the double-log divergence using a ``local divergence conjecture" which, while reasonable, was not proven in their paper\footnote{Previously, an even more singular $(\log \Lambda_y/|q_y|)^5$ divergence was identified in the fundamental large $N$ expansion with no $\epsilon$-deformation~\cite{Holder2015}. However, the authors of Ref.~\cite{Holder2015} were able to remove the divergence by adding an extra term proportional to $q_x^2/q_y^2$ in the boson kinetic energy. Since $q_x^2/q_y^2$ has the same tree-level scaling dimension as $q_y^2$ under patch scaling $q_x \sim q_y^2$, it is natural for $q_x^2/q_y^2$ to be generated by an RG flow. It would be interesting to explore if an analogous workaround can be found for the $(\log \Lambda_y/|q_y|)^2$ divergence in Ref.~\cite{Ye2021}, which has a distinct conceptual origin.}. Due to this unresolved subtlety in the argument, we will temporarily neglect the double-log divergence in this work. If this divergence becomes firmly established in the future (like for the antiferromagnetic quantum critical metal~\cite{Borges2022}), the conclusions that we draw in Section~\ref{sec:transport_expansion} and~\ref{sec:application} still hold over some intermediate energy scales but may break down in the deep IR limit. 

\subsection{Infinitely many diagrams at every order in perturbation theory in the transport regime}\label{subsec:infinitude}

The discussion in Section~\ref{subsec:mross_basic} neglects a fundamental problem: for any correlation function with external momentum $q$ and frequency $\Omega$, the quantum critical regime where $\Omega \ll q$ can have scaling behaviors that are very different from the transport regime $\Omega \gg q$. In other words, the low frequency and low momentum limits do not commute. This phenomenon can be illustrated most simply for the one-loop boson self energy
\begin{equation}
    \Pi_{\rm 1-loop}(\bs{q}, i\Omega) \sim \begin{cases}
        \gamma \frac{|\Omega|}{|\bs{q}|} & \Omega \ll q \\ \Pi_0 & \Omega \gg q 
    \end{cases} \,.
\end{equation}
In the quantum critical regime, the boson self energy correctly reproduces the divergence of the order parameter susceptibility as $m^2 \rightarrow 0$ and also predicts the dynamical critical exponent $z = 2+\epsilon$ when combined with the bare non-analytic kinetic term $|\bs{q}|^{1+\epsilon}$. In the transport regime, the boson self energy is insensitive to the quantum critical scaling behavior and instead approaches a finite positive constant $\Pi_0$.\footnote{The value of $\Pi_0$ does not matter for the present discussion and we defer a precise definition to \eqref{eq:Pi0_def} in Section~\ref{sec:application}.}

\begin{figure}[ht]
    \centering
    \includegraphics[width = \linewidth]{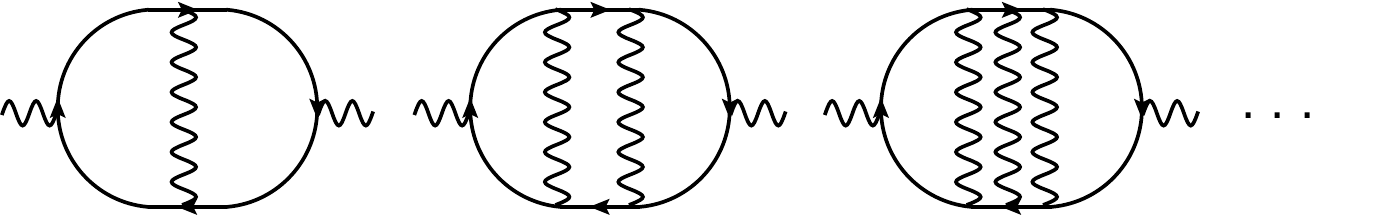}
    \caption{While combinatorial scaling predicts that a ladder diagram with $n$ rungs is $\mathcal{O}(N^{-n})$, an explicit calculation shows that integrals over internal momenta carry additional factors of $N$ and all of the ladder diagrams are in fact $\mathcal{O}(1)$ in the transport regime.}
    \label{fig:ladder}
\end{figure}

This simple observation opens up the possibility that at each order in $1/N$, the set of diagrams contributing in the quantum critical regime can be drastically different from the set of diagrams contributing in the transport regime. We now show that this is already the case even at leading order in $1/N$ by considering ladder diagrams for the boson self energy as shown in Fig.~\ref{fig:ladder}. 

By naive combinatorial scaling, we associate a factor of $N$ to each fermion loop and a factor of $\frac{1}{\sqrt{N}}$ to each boson-fermion
vertex. A ladder with $n$ rungs has $2n+2$ vertices and a single fermion loop. Therefore, one might expect that a ladder with $n$ rungs scales as $N^{-n}$ in the large $N$ limit. However, if we explicitly evaluate the 1-rung diagram in the transport regime, we find a structure~\cite{Kim1994}
\begin{equation}
    \Pi_{\rm 1-rung}(\bs{q}, \Omega) \sim \frac{1}{N\epsilon} \pi(\epsilon N, \bs{q}, \Omega)
\end{equation}
where $\pi$ is a regular function in the large $N$ limit. Since $N \epsilon$ is fixed, this formally $\mathcal{O}(N^{-1})$ contribution is in fact $\mathcal{O}(1)$ and the combinatorial scaling breaks down.

The origin of this breakdown is that integrals over internal momenta can carry anomalous factors of $N$ in the large $N$ limit. By manipulations detailed in~\cite{Kim1994}, one can show that the 1-rung diagram is proportional to the following integral\footnote{Technically, this scaling estimate should have a prefactor which is proportional to powers of the Yukawa interaction form factor $g^a(\bs{k}, \bs{q})$. So long as $g^a(\bs{k}, \bs{q})$ approaches a nonzero limit as $\bs{q} \rightarrow 0$ for most $\bs{k}$ on the Fermi surface (i.e. all but a measure-zero set), this scaling is valid. This constraint on $g^a(\bs{k}, \bs{q})$ is satisfied by most metallic quantum critical points of interest and will be assumed throughout the paper.}
\begin{equation}
    \int_0^{\infty} dq \, D(q, \Omega) \sim \int_0^{\infty} \frac{dq}{q^{1+\epsilon} + \gamma \frac{|\Omega|}{q}} \,.
\end{equation}
For all $\epsilon > 0$, the integral is convergent. However, the value of the integral diverges as $\frac{1}{\epsilon} \sim N$ in the large $N$ limit. In fact, one can check that an $n$-rung diagram contains $n$ powers of this kind of loop integral. Therefore, $\Pi_{\rm n-rung}(\bs{q}=0,\Omega)$ is $\mathcal{O}(1)$ for all values of $n$ and we need to \textit{resum an infinite number of diagrams already at leading order in the double expansion}. 

\section{Extension of the controlled expansion to the transport regime}\label{sec:transport_expansion}

The infinitude of diagrams identified in Section~\ref{subsec:infinitude} seems hopeless at first. The goal of this section is to show that this is not the case and provide a tractable method to calculate their total contribution order by order in the double expansion. In Section~\ref{subsec:geometric_ph}, we first define a class of particle-hole irreducible diagrams $K$ such that the boson self energy can be written as a geometric series $\frac{1}{1-K}$.\footnote{This notation is schematic. The precise definition of this geometric series will be given in Section~\ref{subsec:geometric_ph}.} We then show in Section~\ref{subsec:surgery_suture} that the kernel $K$ only receives contributions from a finite number of diagrams at each order in $1/N$. Moreover, every diagram in the transport regime can be generated by sewing together diagrams in the quantum critical regime in a well-defined way. In Section~\ref{subsec:surgery_suture_conductivity}, we relate the diagrammatic expansion for the boson self energy to the current-current correlation function which controls electrical transport. For readers who are only interested in the main results, the recipe for diagrammatic constructions is given in Section~\ref{subsubsec:recipe} and the formula for conductivity is summarized in Fig.~\ref{fig:G_JJ_decomp} and \eqref{eq:G_JJ_generalform}. With this outline in mind, let us dive in. 

\subsection{Boson self energy as a geometric sum of the particle-hole irreducible kernel}\label{subsec:geometric_ph}

In complete generality, the boson self energy $\Pi(\bs{q} = 0, \Omega)$ consists of all diagrams with two external boson legs that are irreducible with respect to cutting a single internal boson propagator. In each diagram, we assume that the IR singular contributions come from the integration region where all internal propagators live in the quantum critical regime. When this is the case, we can first resum all the quantum critical self energy corrections in internal propagators. 
After the resummation step, the only diagrams we need to evaluate are \textit{skeleton diagrams} with no internal self energy subdiagram. 

\begin{figure}
    \centering
    \begin{subfigure}{.6\linewidth}
    \includegraphics[width=\linewidth]{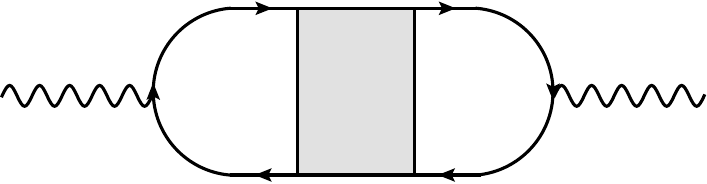}
    \end{subfigure}
    
    \vspace{0.6cm}
    \centering
    \begin{subfigure}{\linewidth}
    \includegraphics[width=\linewidth]{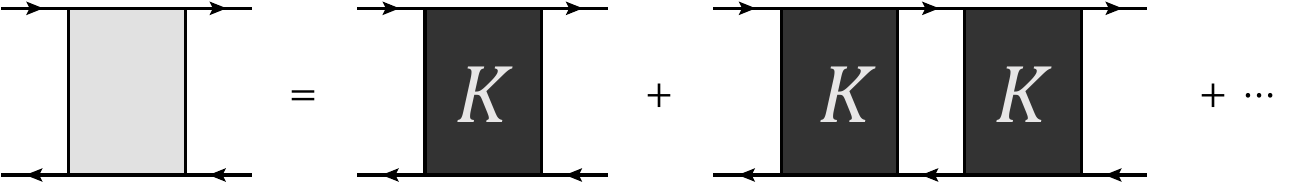}
    \end{subfigure}
    \caption{The one particle irreducible boson self energy can be decomposed into a 4-pt fermion vertex multiplied by four fermion propagators. The 4-pt fermion vertex (the grey box) is a geometric sum of the particle-hole irreducible kernel $K$.}
    \label{fig:ph_irreducible}
\end{figure}

We now proceed to organize the remaining vertex corrections encoded in the \textit{skeleton diagrams}. The key observation is that $\Pi(\bs{q} =0, \Omega)$ can be recast as a geometric series of particle-hole irreducible kernel $K$, as shown in Fig.~\ref{fig:ph_irreducible}.
Intuitively, the particle-hole irreducible kernel $K$ consists of diagrams that cannot be factorized by cutting a particle propagator on the top and a hole propagator on the bottom. In the rest of this section, we precisely define the meaning of these terms and show that the geometric series is a complete expansion. 

\begin{figure}
    \centering
    \begin{subfigure}{0.35\linewidth}
    \includegraphics[width=\linewidth]{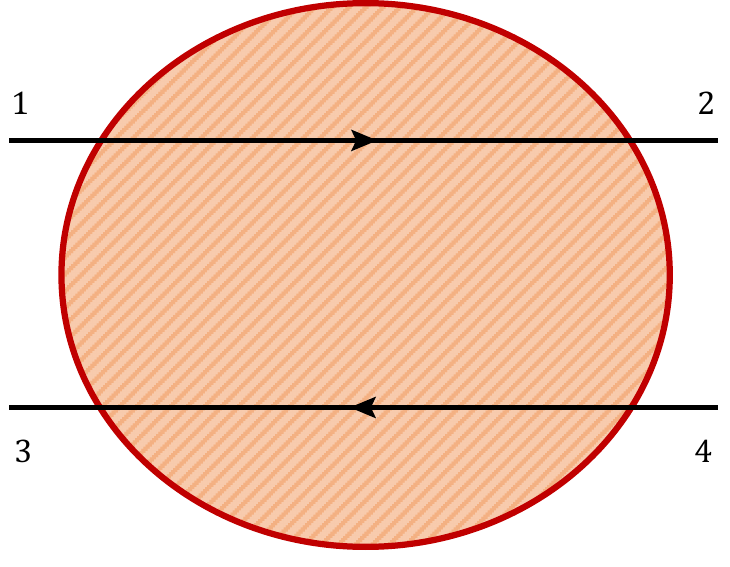}
    \caption{horizontal fermion four-point vertex}
    \end{subfigure}
    \begin{subfigure}{0.36\linewidth}
    \includegraphics[width=\linewidth]{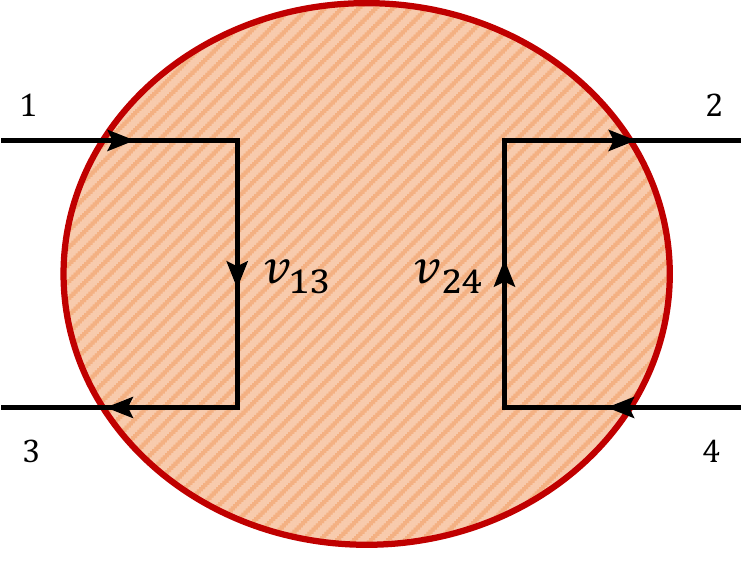}
    \caption{vertical fermion four-point vertex}
    \end{subfigure}
    \caption{The canonical representation of horizontal and vertical fermion four-point vertex. The shaded region can contain arbitrarily many internal propagators and loops. Each solid line in the shaded regions do not represent a single fermion propagator but rather a series of fermion propagators running in the same direction (horizontal/vertical). }
    \label{fig:def_4ptvertex}
\end{figure}

Consider a generic four-fermion vertex diagram with two particle legs labeled as 1, 2 and two hole legs labeled as 3, 4. This vertex is shown in Fig.~\ref{fig:def_4ptvertex}, with the orange shaded region enclosing an arbitrary number of internal propagators. Since the interaction vertex couples a particle-hole operator to a boson, fermion lines in the diagram cannot terminate. Therefore, the four legs 1, 2, 3, 4 must be pairwise connected via a sequence of internal fermion propagators. Moreover, since every interaction vertex must have one incoming fermion line and one outgoing fermion line, 1 must be connected to 2 or 3, but not 4. When 1 connects to 2 and 3 connects to 4, we can always arrange so that both connections consist of completely horizontal internal fermion propagators. This defines the canonical representation of a \textbf{horizontal four-point vertex}. When 1 connects to 3, we can always arrange so that the connection between 1 and 3 goes through a series of horizontal fermion propagators starting at 1, turns towards a series of vertical fermion propagators labeled as $v_{13}$, and finally turns again to a series of horizontal fermion propagators running towards 3. A similar arrangement can be made for the connection between 2 and 4. This procedure defines the canonical representation of a \textbf{vertical four-point vertex}. These two cases are represented in Fig.~\ref{fig:def_4ptvertex}.

We now think about factorizations. Suppose we start out with a horizontal four-point vertex. Then we can apply a recursive procedure, where in each step, we look for a virtual string that cuts across one particle propagator on the 1-2 connection and one hole propagator on the 3-4 connection and nothing else. If such a line can be found, we say that the horizontal diagram is \textbf{p-h reducible}. Otherwise, it is \textbf{p-h irreducible}. Since each cut reduces the number of internal lines in the original diagram, this process must terminate in finitely many steps, in which case we are left with a finite product of \textbf{horizontal p-h irreducible diagrams}. 

\begin{figure}[ht]
    \centering
    \begin{subfigure}{0.44\linewidth}
    \includegraphics[width=0.85\linewidth]{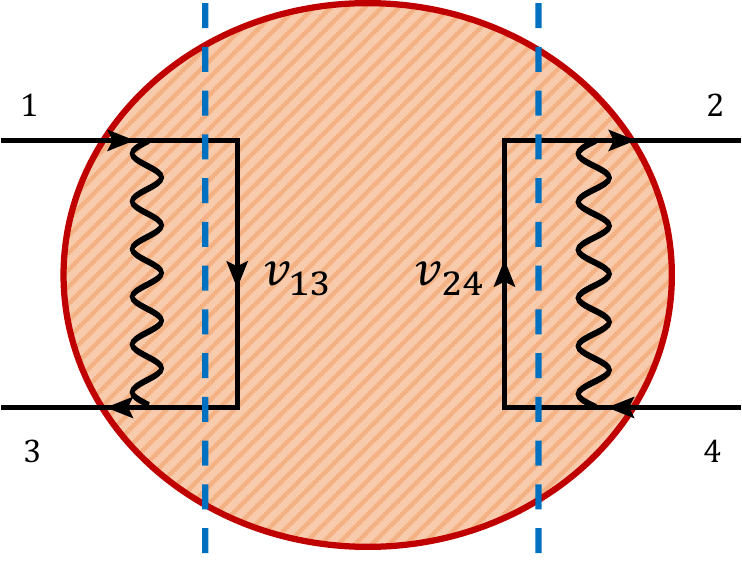}
    \caption{A non-minimal vertical four-point vertex.}
    \end{subfigure}
    \begin{subfigure}{0.43\linewidth}
    \includegraphics[width=0.85\linewidth]{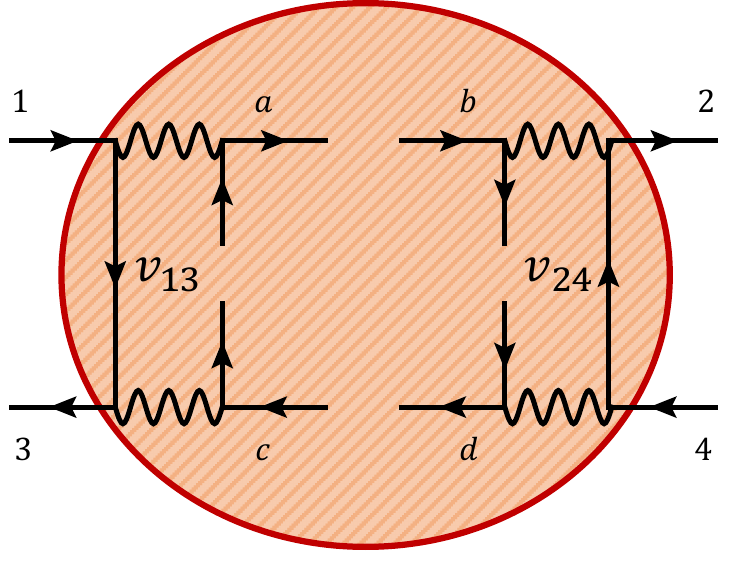}
    \caption{A minimal vertical four-point vertex.}
    \end{subfigure}
    \caption{A vertical four-point vertex is minimal/non-minimal if it cannot/can be factorized by a vertical cut on the left of $v_{13}$ or on the right of $v_{24}$. We illustrate both cases in this figure.}
    \label{fig:def_4ptvertex_minimalvertical}
\end{figure}

Suppose instead that we start out with a vertical four-point fermion vertex, where $v_{13}$ and $v_{24}$ are vertical sections of the diagram that contain a series of consecutive fermion propagators, interrupted by outgoing boson propagators that are not explicitly represented in Fig.~\ref{fig:def_4ptvertex}. If we apply a vertical cut on the immediate left of $v_{13}$ and on the immediate right of $v_{24}$, then the original diagram factorizes into a product of two horizontal four-point vertices and one vertical four-point vertex (see Fig.~\ref{fig:def_4ptvertex_minimalvertical}(a) for an example of such a factorization). Following the procedure in the previous paragraph, the horizontal four-point vertex diagrams can be further factorized into products of horizontal p-h irreducible diagrams. Therefore, we need only analyze the factorization of a vertical four-point vertex diagram where the external legs are immediately next to $v_{13}, v_{24}$ (i.e. it cannot be further reduced from the left of $v_{13}$ or from the right of $v_{24}$). 

Due to the Yukawa interaction structure, the four corners where vertical and horizontal fermion lines connect (which we call NW, NE, SW, SE) are each connected to one boson propagator as shown in Fig.~\ref{fig:def_4ptvertex_minimalvertical}(b). We canonically choose these boson propagators to be horizontal. The other end of each boson line must be connected to one incoming and one outgoing fermion leg. We define a/d as the outgoing fermion leg at the NW/SE vertex, and b/c as the incoming fermion leg at the NE/SW vertex. With these definitions, it is natural to refer to a, b as particle lines and c, d as hole lines. Now we say that the \textbf{vertical diagram is p-h reducible} if a string can be drawn that cuts only through one of a, b and one of c, d. Otherwise it is \textbf{p-h irreducible}. After making one such cut, the diagram factorizes into a product of two smaller vertical four-point vertex diagrams. This recursive process again terminates in finitely many steps. Therefore, we arrive at the final conclusion:

\noindent \textbf{Claim 1}: \textit{Every four-point vertex can be reduced to a finite product of diagrams, each one of which is either horizontal p-h irreducible or vertical p-h irreducible}. 

From here on, we will refer to the sum of all p-h irreducible four-point fermion vertex diagrams (horizontal and vertical) as $K$. Since the kernel $K$ has four external legs, we regard it as a linear operator acting on two external legs with average momentum $\bs{k}$, average frequency $\omega$, relative momentum $\bs{q}$, and relative frequency $\Omega$. These functions of four variables span a vector space. Using a braket notation, we can represent the boson vertex as $\ket{g}$ such that
\begin{equation}
    \bra{\bs{k}, \omega, \bs{q}, \Omega}\ket{g_{\bs{\bar q}, \bar \Omega}} = g(\bs{k}, \bs{q}) \,\delta^2(\bs{q} - \bs{\bar q})\, \delta(\Omega - \bar \Omega) \,.
\end{equation}
The implication of Claim 1 is that the full boson self energy can be rewritten as a geometric series in $K$:
\begin{equation}
    \Pi(\bs{q}, \Omega) = \bra{g_{\bs{q},\Omega}} W_G (1 - K)^{-1} \ket{g_{\bs{q},\Omega}} \,,
\end{equation}
where $W_G$ is the kernel that corresponds to two parallel-running fermion propagators
\begin{equation}
    \begin{aligned}
    \bra{\bs{k},\bs{q}, \omega, \Omega}W_G\ket{\bs{k}',\bs{q}',\omega', \Omega} &= \delta^2(\bs{k} - \bs{k}') \,\delta^2(\bs{q} - \bs{q'})\, \delta(\omega - \omega') \,\delta(\Omega - \Omega') \\
    &\cdot G(\bs{k} + \bs{q}/2, i\omega + i\Omega/2) \, G(\bs{k} - \bs{q}/2, i\omega-i\Omega/2) \,.
    \end{aligned}
\end{equation}
Having defined this compact braket notation, we proceed to find a controlled expansion for the irreducible kernel $K$. In general, the matrix elements of the irreducible kernel $\bra{v_1}K\ket{v_2}$ can depend in complicated ways on the choice of $v_1, v_2$. However, since possible anomalous factors of $N$ come from the internal loop integrals over boson momenta in $K$, the $N$-counting of the matrix elements is independent of $v_1, v_2$. Therefore, as far as $N$-counting is concerned, it suffices to contract the four dangling legs in $K$ with two external boson vertices (which corresponds to choosing $\ket{v_1} = W_G \ket{g_{\bs{q}, \Omega}}$ and $\ket{v_2} = \ket{g_{\bs{q}, \Omega}}$). For convenience, we will slightly abuse notations and refer to this contraction $\bra{v_1} K \ket{v_2}$ as $K$. It should be understood that this contraction captures the $N$-scaling of the original kernel $K$ but not its detailed functional form.

\subsection{A bridge between transport and quantum critical diagrams: surgery and suture}\label{subsec:surgery_suture}

In the previous section, we gave a precise definition of $K$, which is the sum of all p-h irreducible diagrams for the fermion four-point vertex. Given this preparation, we are now ready to present a classification of all diagrams in $K$ in the transport regime. 

The classification scheme hinges upon a basic observation: independent of the choice of external momentum/frequency, the dominant singular contributions to every loop diagram in the infrared limit arise from the integration region where all \textit{internal boson propagators} are in the quantum critical regime $\Omega \ll q$. To see that, we recall the large $N$ solution for the Euclidean boson propagator with $q \ll k_F$ and arbitrary $\Omega$ (up to some microscopic energy scale $E_{\rm UV}$):
\begin{equation}
    D(q, i\Omega) \sim \begin{cases} \Omega^{-2} & E_{\rm UV} \sim \Omega \gg q \\ \Pi_0^{-1} & E_{\rm UV} \gg \Omega \gg q \\ \left(q^{1+\epsilon} + \frac{|\Omega|}{q} \right)^{-1} & \Omega \ll q \end{cases} \,.
\end{equation}
For every internal boson propagator, if we fix a small $q \ll k_F$ and integrate over $\Omega$, the gapless virtual processes at $\Omega \ll q$ strongly enhance $D(q, i\Omega)$ and provide singular contributions, while the gapped virtual processes with $\Omega \gg q$ give a subleading correction that depends on the UV cutoff $E_{\rm UV}$. Therefore, for the purpose of identifying the dominant contributions to each diagram and classifying their $N$-scaling, it suffices to restrict all internal boson legs to the quantum critical regime.

\begin{figure}[ht]
    \centering
    \includegraphics[width = \linewidth]{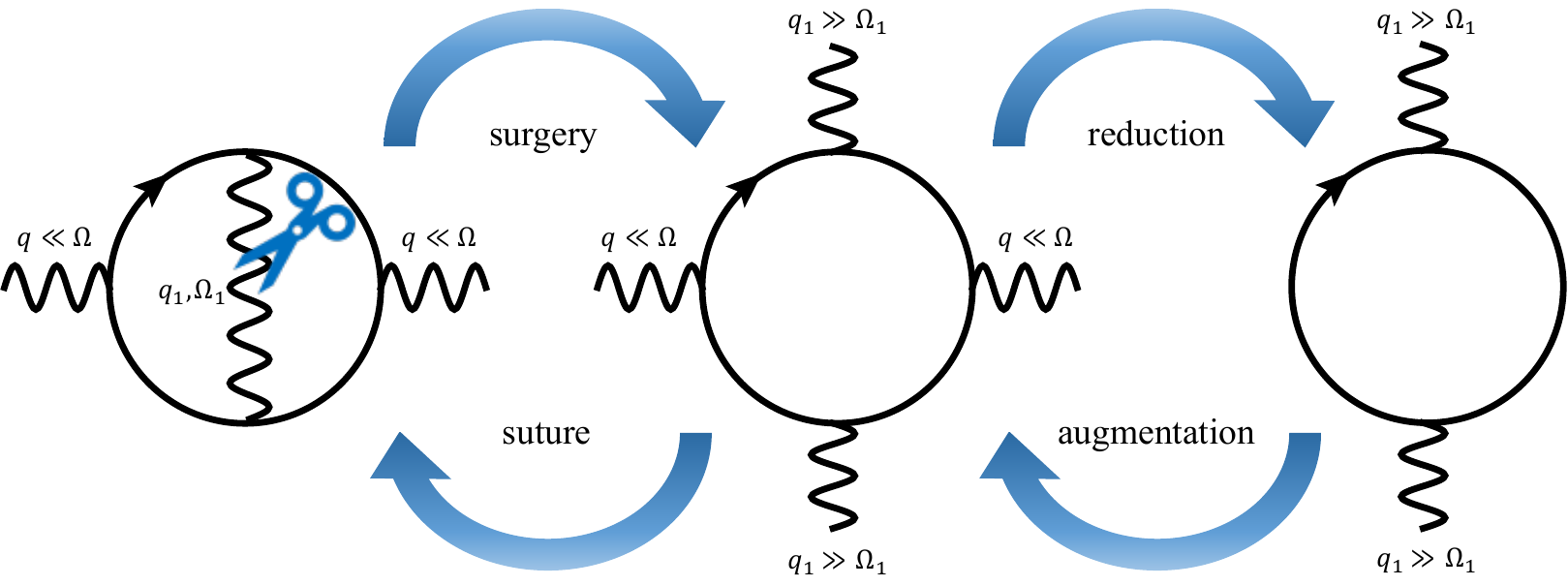}
    \caption{A pictorial recipe for generating a quantum critical diagram from a transport diagram (and vice versa). Reading from the left, the surgery step cuts one internal boson propagator open, and the reduction step removes the two external boson legs. The augmentation and suture steps run in reverse. The key observation is that the $N$-scaling of the transport diagram on the left can be determined by the $N$-scaling of the quantum critical diagram on the right.}
    \label{fig:surgery_steps}
\end{figure}

With this preliminary observation in mind, we can give a broad summary of the classification. The key idea is that every transport diagram becomes a (possibly reducible) quantum critical diagram after we cut one of the internal boson propagators and remove the original external boson legs. Since there are many internal boson propagators in general, this correspondence is not one-to-one. However, if we can enumerate the set $\mathcal{S}$ of \textit{all possible} quantum critical diagrams that can arise from cutting \textit{any internal propagator} in skeleton transport diagrams, we can then run the surgery in reverse and obtain the set of all transport diagrams by sewing together the external boson legs of every quantum critical diagram in the set $\mathcal{S}$ and then removing the redundancies. As a concrete example, consider the Maki-Thompson diagram in the transport limit, as represented on the left side of Fig.~\ref{fig:surgery_steps}. Reading Fig.~\ref{fig:surgery_steps} from left to right, the surgery step involves cutting the internal boson propagator with momentum $q_1$ and frequency $\Omega_1 \ll q_1$. The reduction step then removes the two external boson lines with momentum $q$ and frequency $\Omega \gg q$. In the end, we are left with a quantum critical diagram which is just a one-loop fermion bubble. Reading Fig.~\ref{fig:surgery_steps} from right to left, the augmentation and suture steps are inverses of the reduction and surgery steps. 

What we will show is that the $N$-scaling of every transport diagram is upper bounded by the $N$-scaling of the quantum critical diagram obtained by ``surgery + reduction". Since all quantum critical diagrams follow combinatorial scaling, and since there are only finitely many quantum critical diagrams at each order in $1/N$, the bridge between transport and quantum critical diagrams immediately implies a controlled and tractable expansion for the kernel $K$ in the transport regime. 

\subsubsection{From transport diagrams to quantum critical diagrams and back again}\label{subsubsec:surgery_suture_explained}

\noindent \textbf{Step 1: surgery and reduction}

Consider an arbitrary diagram $K_i$ in the transport regime that contributes to the p-h irreducible kernel $K$. With the exception of the trivial identity diagram, every $K_i$ has at least one internal boson line carrying momentum $\bs{q}_1$ and frequency $\Omega_1$. When we cut this internal boson line, we obtain a new diagram with four external legs (for concreteness, one can again refer to the middle part of Fig.~\ref{fig:surgery_steps}). Two of these legs carry momentum $\bs{q}_1$ and frequency $\Omega_1$ which are in the quantum critical regime $\Omega_1 \ll |\bs{q}_1|$. The other two legs carry momentum $\bs{q}$ and frequency $\Omega$ which are in the transport regime $\Omega \gg |\bs{q}|$. When $\bs{q} \rightarrow 0$, momentum is conserved at the transport legs. Therefore, as far as momentum integrals are concerned, one can further delete the two external legs in the transport regime without changing the $N$ scaling of loop integrals. But because we lose two vertices in this process, we must multiply the original diagram by a factor of $N$. 

\noindent \textbf{Step 2: augmentation and suture}

After the surgery and reduction procedure, we obtain an amputated diagram $\Pi_{\rm amp, i}(\bs{q}_1, \Omega_1)$ which can be regarded as a boson self energy diagram in the quantum critical regime. By running the procedure in reverse, we can relate the $N$-scaling of the original diagram $K_i$ to the amputated diagram by a loop integral 
\begin{equation}\label{eq:suture_loop}
    K_i \sim  \frac{1}{N}\, \int_{\bs{q}_1, \Omega_1} D(\bs{q}_1, \Omega_1) \,\Pi_{\rm amp, i}(\bs{q}_1, \Omega_1) \,,
\end{equation} 
where $D$ is the renormalized boson propagator, and $1/N$ comes from the augmentation step which brings in two extra vertices. 

Since $\bs{q}_1, \Omega_1$ started their lives as the momentum/frequency of an internal boson propagator, $\Pi_{\rm amp, i}(\bs{q}_1, \Omega_1)$ should be evaluated in the quantum critical regime, where combinatorial counting is valid. Therefore, we can write
\begin{equation}
    \Pi_{\rm amp, i}(\bs{q}_1, \Omega_1) = N^{-n_i} \, \pi_{\rm amp, i}(\bs{q}_1, \Omega_1) \,,
\end{equation}
where $N^{-n_i}$ is the combinatorial scaling of $\Pi_{\rm amp,i}$ and $\pi_{\rm amp, i}$ is an $\mathcal{O}(1)$ function of its arguments. The final task is to determine the scaling of the integral over $\bs{q}_1, \Omega_1$. But since the double expansion is controlled in the quantum critical regime, higher order corrections to $\Pi(\Omega_1 \ll q_1)$ must be comparable/subleading relative to the leading Landau damping term $|\Omega_1|/|\bs{q}_1|$ under dynamical critical scaling. In particular, this means that 
\begin{equation}
    \pi_{\rm amp, i}(\bs{q}_1, \Omega_1) \propto c_0 \frac{|\Omega_1|}{|\bs{q}_1|} + \text{subleading} \,,
\end{equation}
where the subleading terms are suppressed by additional powers of $|\Omega_1|/E_F$ and/or $|\Omega_1|/|\bs{q}_1|$. Plugging $\pi_{\rm amp,i}$ back into the integral \eqref{eq:suture_loop}, we find 
\begin{equation}
    \begin{aligned}
    K_i &\sim N^{-n_i-1} \int_{\bs{q}_1, \Omega_1} \frac{1}{|\bs{q}_1|^{1+\epsilon} + \frac{|\Omega_1|}{|\bs{q}_1|} + \mathcal{O}(1/N)} \cdot \pi_{\rm amp,i}(\bs{q}_1, \Omega_1, \Omega) \\
    &\sim N^{-n_i-1} \int_{\bs{q}_1, \Omega_1} \frac{1}{|\bs{q}_1|^{1+\epsilon} + \frac{|\Omega_1|}{|\bs{q}_1|} + \mathcal{O}(1/N)} \cdot \left(c_0 \frac{|\Omega_1|}{|\bs{q}_1|} + \text{subleading}\right) \,.
    \end{aligned}
\end{equation}
The integral over frequency is superficially divergent. However, this divergence is innocuous because it can be regulated by UV frequency cutoffs that do not depend explicitly on $\epsilon$. On the other hand, potential divergences in the momentum integral would be more concerning as they are related to scattering processes along the Fermi surface mediated by the gapless boson, which can have large momentum but low energy. Fortunately, the presence of a positive $\epsilon$ guarantees the convergence of momentum integrals. If $c_0 \neq 0$, then the integral over $\bs{q}_1$ scales as $1/\epsilon$ at small $\epsilon$. If $c_0 = 0$, then the integral is $\mathcal{O}(1)$ in the small $\epsilon$ limit. Therefore, we conclude that there is an upper bound on $K_i$
\begin{equation}
    K_i \lesssim N^{-n_i-1} \cdot \frac{1}{\epsilon} \sim N^{-n_i} \,.
\end{equation}
This upper bound tells us that every transport diagram that contributes at $\mathcal{O}(N^{-n_i})$ can be related (not uniquely) to a quantum critical diagram $\Pi_{\rm amp, i}$ which scales as $\mathcal{O}(N^{-n_i})$ according to fundamental large $N$ counting. 

\noindent \textbf{Step 3: classification of all possible $\Pi_{\rm amp, i}$}

The final step is to devise a simple method to classify and enumerate the set of all quantum critical diagrams that can be obtained from a transport diagram by the ``surgery + reduction" procedure defined in Step 2. For this purpose, it is useful to define a generalized notion of diagrammatic reducibility as follows.

\noindent \textbf{Definition}: A diagram is \textbf{boson-n-reducible} if it can be factorized into $n+1$ disconnected subdiagrams by cutting $n$ internal boson lines. 

With this definition in mind, we can establish two claims (see Appendix~\ref{app:proofs} for detailed proofs).

\noindent \textbf{Claim 2}: $\Pi_{\rm amp,i}$ is not boson-$n$-reducible for any $n > 1$.

\noindent \textbf{Claim 3}: $\Pi_{\rm amp,i}$ does not contain a fermion self energy (fSE) subdiagram. 

Based on these claims, we arrive at the important conclusion: \textit{the set of all possible $\Pi_{\rm amp,i}$ is the set of all boson self energy diagrams that are p-h irreducible, free from fSE subdiagrams, and not boson-$n$-reducible for any $n > 1$}. 

\subsubsection{Explicit recipe for constructing all transport diagrams at each order in the controlled expansion}\label{subsubsec:recipe}

The classification in Step 3 immediately implies a simple procedure that generates all possible transport diagrams in $K$ at any order in the double expansion. Suppose we want to calculate the $\mathcal{O}(N^{-n})$ contribution to the boson self energy in the transport regime. Then we first enumerate all boson self energy diagrams that are p-h irreducible, free from fermion self energy (fSE) subdiagrams, not boson-$m$-reducible for $m > 1$, and $\mathcal{O}(N^{-n})$ under fundamental large $N$ counting. Label these diagrams by an index $i$. For each $i$, we then sew the two external boson legs of diagram $i$ together. Now enumerate all possible ways to cut two internal fermion lines and obtain a particle-hole kernel with four external fermion legs. The non-redundant kernels generated from this procedure can be labeled as $K_{i,s}$.

The total contribution to the particle-hole irreducible kernel at $\mathcal{O}(N^{-n})$ is therefore $K^{(n)} = \sum_{s, n_i = n} K_{i,s}$.\footnote{Remember that some diagrams with $n_i = n$ might actually be $\ll N^{-n}$. But the important claim is that diagrams with $n_i > n$ are always $\ll N^{-n}$.} The boson self energy in the transport regime can be obtained via a geometric sum of the kernel 
\begin{equation}
    \Pi(\bs{q}=0,\Omega) = \bra{g_{\bs{q}=0,\Omega}} W_G \frac{1}{1 - \sum_n K^{(n)}} \ket{g_{\bs{q}=0,\Omega}}
\end{equation}
where $\ket{g_{\bs{q}, \Omega}}, W_G$ are defined in Section~\ref{subsec:geometric_ph}. This explicit recipe is the main technical result of this paper. Note that although the expansion has been developed for the boson self energy, the large $N$ counting structure in fact generalizes to two-point correlation functions of \textit{any fermion bilinear operator}. In the next section, we use this observation to derive a diagrammatic expansion for the electrical conductivity. 

\subsection{Relating the boson self energy to the optical conductivity}\label{subsec:surgery_suture_conductivity}

The electrical conductivity associated with a $U(1)$ conserved current $\bs{J}$ is related to a two-point function of $\bs{J}$ via the Kubo formula
\begin{equation}
    \sigma_{ij}(\Omega) = \frac{G_{J_i J_j}(\bs{q}=0, \Omega)}{i\Omega} \,, \quad J_i(\bs{q}) = \int \frac{d^2 \bs{k}}{(2\pi)^2} v^i_F(\bs{k}) \sum_{i=1}^N f^{\dagger}_i(\bs{k} + \bs{q}/2) f_i(\bs{k} - \bs{q}/2) \,.
\end{equation}
The computation of $G_{J_i J_j}$ requires evaluating all possible Feynman diagrams that end on two external current vertices. Pictorially, to distinguish from the Yukawa interaction vertex, we represent the current vertex by a blue wavy line connected to a pair of incoming/outgoing fermion lines. Unlike for the boson self energy, the Feynman diagrams for $G_{J_iJ_j}$ need not be irreducible with respect to single boson line cuts. Therefore, we cannot directly write $G_{J_i J_j}$ as a geometric series of p-h irreducible diagrams. However, this problem can be easily resolved if we decompose $G_{J_iJ_j}$ into two terms-one which contains all one-boson-line irreducible diagrams, and one which contains the rest. 

\begin{figure}[ht]
    \centering
    \includegraphics[width=\linewidth]{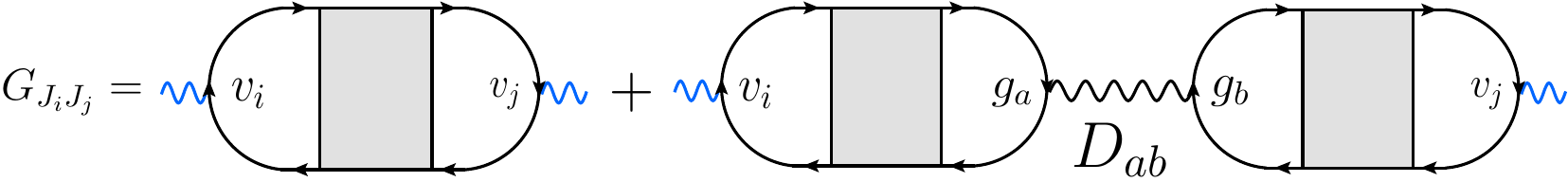}
    \caption{The current-current correlator can be decomposed into two terms. In the first term, two external current vertices are directly connected by the grey box which is the familiar 4-point fermion vertex that we encountered in Section~\ref{subsec:geometric_ph}. The interior of the box is irreducible with respect to single boson line cuts. In the second term, we account for the remaining reducible diagrams by resumming them into an exact boson propagator $D_{ab}$.}
    \label{fig:G_JJ_decomp}
\end{figure}

The first term is simple to represent in terms of the four-point fermion vertex defined in Section~\ref{subsec:geometric_ph}. Following the notation of Section~\ref{subsec:geometric_ph}, we draw the four-point fermion vertex as a grey box and insert it between the two external current vertices (this is the first term in Fig.~\ref{fig:G_JJ_decomp}). For a non-interacting Fermi surface, this is the only term that survives and it gives rise to the Drude weight. The second term contains all diagrams that are reducible by cutting one internal boson line. But any such diagram can always be factorized into a product of one-boson-line irreducible diagrams. The factors that are disconnected from the external current vertices resum into a fully renormalized boson propagator. The factors that are connected to the external current vertices are equivalent to two additional grey box insertions. Therefore, the second term also admits a simple diagrammatic representation as illustrated in Fig.~\ref{fig:G_JJ_decomp}.

Finally, let us recall that the four-point fermion vertex (represented as a grey box) is the geometric series of p-h irreducible diagrams. Using the braket notation introduced in Section~\ref{subsec:geometric_ph}, we can therefore translate Fig.~\ref{fig:G_JJ_decomp} into a compact equation
\begin{equation}\label{eq:G_JJ_generalform}
    N^{-1} G_{J_i J_j} = \bra{v_i} W_G (1-K)^{-1} \ket{v_j} + \bra{v_i} W_G (1-K)^{-1} \ket{g_a} D_{ab} \bra{g_b} W_G (1-K)^{-1} \ket{v_j} \,.
\end{equation}
Note that with $N$ species of fermions, the conductivity should scale as $N$. The factor of $N^{-1}$ on the LHS guarantees that the RHS is an $\mathcal{O}(1)$ quantity. Once we have a controlled expansion for $K$, the electrical conductivity can be extracted by evaluating the inner products in \eqref{eq:G_JJ_generalform}. In practice, this procedure is challenging because we need to invert an infinite-dimensional operator $1-K$. However, as we will see in Section~\ref{sec:application}, it is often the case that the spectrum of $1-K$ contains a set of near-zero eigenvalues which dominate the IR behavior of the conductivity. This simplification allows us to obtain explicit formulae for the leading frequency dependence of $\sigma(\bs{q}=0,\Omega)$.

\section{Applications}\label{sec:application}

We now apply the formalism developed in Section~\ref{sec:transport_expansion} to two concrete physical examples. The first example is a Fermi surface coupled to a single scalar boson order parameter with a form factor $g(\bs{k})$. We will find that the boson self energy $\Pi(\bs{q} = 0, \Omega)$ approaches a finite constant in the IR limit. This constant self energy in turn gives rise to a $g(\bs{k})$-dependent correction to the free-fermion Drude weight. Both of these results are fully consistent with the non-perturbative anomaly arguments in~\cite{shi2022_gifts,shi2023_loop}. We also calculate incoherent corrections to the optical conductivity and find agreement with a recent analysis in the Yukawa-SYK expansion~\cite{guo2023migdal,guo2023fluctuation}. 

\subsection{Spinless Hertz-Millis model with a single critical boson}\label{subsec:app_1boson}

\begin{figure}[ht]
    \centering
    \includegraphics[width = 0.35 \textwidth]{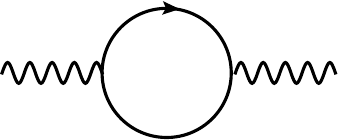}
    \includegraphics[width = 0.57 \textwidth]{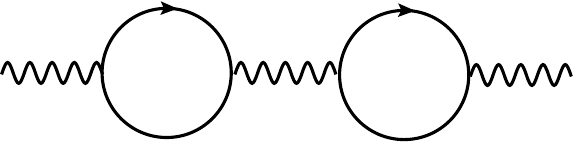}
    \caption{The two amputated diagrams that contribute to $\Pi_{\rm amp}$ to leading order in the double expansion}
    \label{fig:mross_leading}
\end{figure}

Let us recall the Hertz-Millis action for a scalar boson order parameter field within the double expansion
\begin{equation}
    \begin{aligned}
        S &= S_{f} + S_{\phi} + S_{\rm int} \,,\\
        S_{f} &= - \sum_{i=1}^N \int_{\bs{k}, \omega} f^{\dagger}_i(\bs{k}, \omega) \left[i \omega - \epsilon(\bs{k})\right] f_i(\bs{k}, \omega) \,, \\
        S_{\phi} &= \frac{1}{2} \int_{\bs{q}, \Omega} \phi^*(\bs{q}, \Omega) \left[\lambda \Omega^2 + |\bs{q}|^{1+\epsilon}\right] \phi(\bs{q}, \Omega) \,, \\
        S_{\rm int} &= \frac{1}{\sqrt{N}} \sum_{i=1}^N \int_{\bs{k},\bs{q}, \omega, \Omega} g(\bs{k}) \, f^{\dagger}_i(\bs{k}+\bs{q}/2,\omega+\Omega/2) \,f_i(\bs{k}-\bs{q}/2,\omega-\Omega/2) \,\phi(\bs{q}, \Omega)  \,.
    \end{aligned}
\end{equation}
Our goal is to compute the self energy $\Pi(\bs{q} = 0, \Omega)$ in the limit $\Omega \rightarrow 0$ to leading order in $1/N$ with $\epsilon N$ fixed. To do that, let us follow the recipe in Section~\ref{subsubsec:recipe} and first identify the correct set of diagrams to include in $K$. At leading order in the $1/N$ expansion, the set of diagrams that can contribute to $\Pi_{\rm amp}$ consist of all boson self energy diagrams that are p-h irreducible, free from fSE subdiagrams, not boson-$m$-reducible for $m > 1$ and $\mathcal{O}(N^0)$ under fundamental large $N$ counting. By explicit enumeration, one finds that there are only two diagrams that we need to account for, as shown in Fig.~\ref{fig:mross_leading}. 
Let us label the diagram on the left/right as $\Pi_{\rm amp, 1}/\Pi_{\rm amp, 2}$. The next task is to enumerate all possible ways to perform the ``augmentation + suture" procedure in Fig.~\ref{fig:surgery_steps} on each diagram. 

\begin{figure}[ht]
    \centering
    \includegraphics[width = 0.7\textwidth]{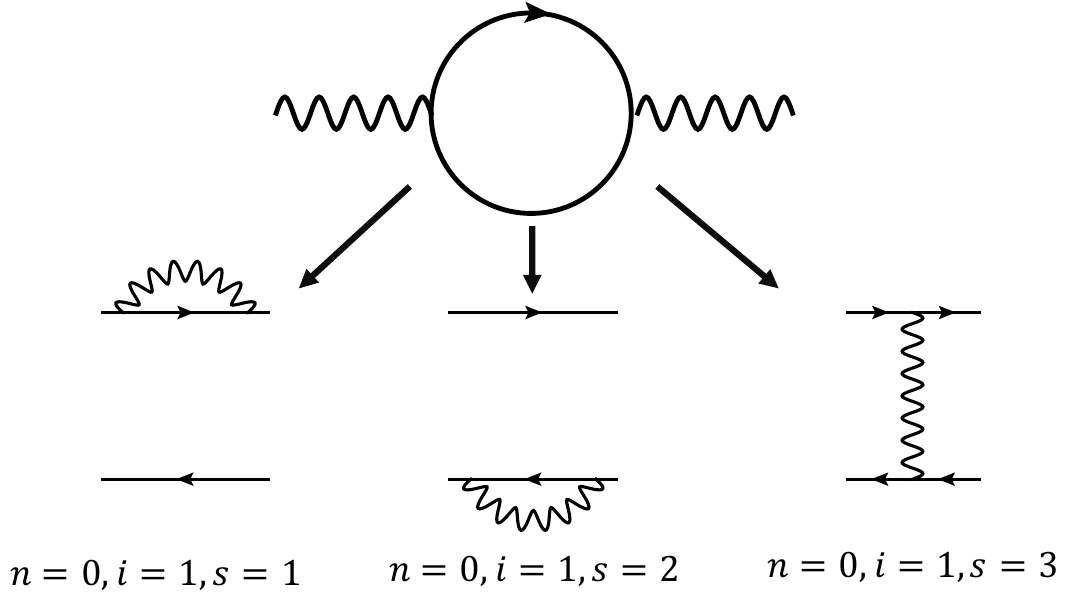}
    \caption{The three diagrams $K^{(n=0)}_{i=1, s= 1, 2, 3}$ that arise from three distinct ways to connect the two external boson legs in the amputated diagram $\Pi_{\rm amp, 1}$.}
    \label{fig:mross_leading_i1suture}
\end{figure}

\begin{figure}[ht]
    \centering
    \includegraphics[width = 0.6\textwidth]{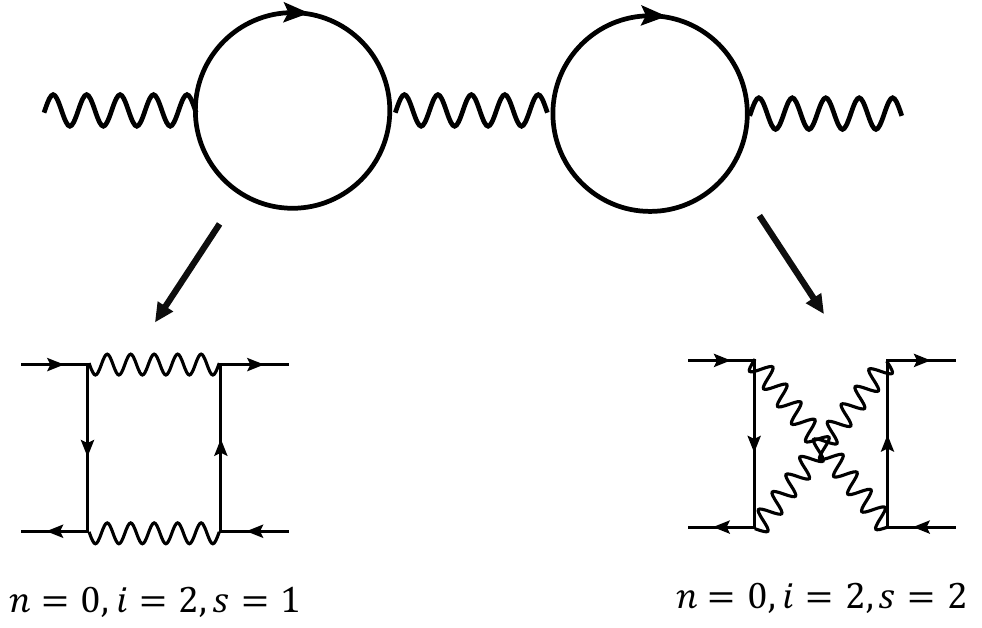}
    \caption{The two diagrams $K^{(n=0)}_{i=2, s = 1, 2}$ that arise from two distinct ways to connect the two external boson legs in the amputated diagram $\Pi_{\rm amp, 2}$.}
    \label{fig:mross_leading_i2suture}
\end{figure}

For $\Pi_{\rm amp, 1}$, there are three ways as shown in Fig.~\ref{fig:mross_leading_i1suture}. The $s = 1$ and $s=2$ diagrams represent particle-hole kernels with an electron self energy correction on the particle/hole line. These diagrams are not admissible because only skeleton diagrams can contribute to $K$. The $s = 3$ diagram is the familiar Maki-Thompson diagram, which is admissible. 
For $\Pi_{\rm amp, 2}$, there are two ways to perform augmentation + suture as shown in Fig.~\ref{fig:mross_leading_i2suture}. Both $s = 1$ and $s=2$ are admissible p-h irreducible diagrams. Together, they constitute the Aslamazov-Larkin corrections to the boson self energy. 

Following conventional notations, we refer to $K^{(n=0)}_{i=1, s = 3}$ as $K_{\rm MT}$ and $K^{(n=0)}_{i=2, s = 1} + K^{(n=0)}_{i=2, s = 2}$ as $K_{\rm AL}$, where MT stands for Maki-Thompson and AL stands for Aslamazov-Larkin. In terms of these kernels, the $\mathcal{O}(N^0)$ result for the self energy reduces to
\begin{equation}\label{eq:HM_1boson_Pi}
    \Pi^{(n=0)}(\bs{q}=0,\Omega) = \bra{g_{\bs{q}=0,\Omega}} W_G (1 - K_{\rm MT} - K_{\rm AL})^{-1} \ket{g_{\bs{q}=0,\Omega}} \,.
\end{equation}
The set of diagrams that need to be resummed in the above expression is identical to the set of diagrams appearing in the Gaussian effective theory for fermion bilinears in the Yukawa-SYK large $N$ model. Therefore, we can directly import results obtained for the Yukawa-SYK model~\cite{guo2022_largeN,shi2023_loop,guo2023migdal,guo2023fluctuation} (allowing for a possible renormalization of the dynamical critical exponent $z$)
\begin{equation}\label{eq:Pi_answerSec4.1}
    \Pi(\bs{q}=0, \Omega) = \Pi_0 + \begin{cases} c_1 \Omega^{(8-z)/z} & \text{convex FS} \\ c_2 \Omega^{(4-z)/z} & \text{concave FS} \end{cases} \,. 
\end{equation}
For a general Fermi surface parametrized by Fermi momentum $\bs{k}_F(\theta)$, the constant part $\Pi_0$ is related to an angular integral over the Fermi surface 
\begin{equation}\label{eq:Pi0_def}
    \Pi_0 = \Tr_{\theta} g^2 \,, \quad \Tr_{\theta} f g = \int d \theta \frac{\left|\partial_{\theta} \bs{k}_F(\theta)\right|}{(2\pi)^2} \frac{f(\theta) g(\theta)}{v_F(\theta)} \,,
\end{equation}
which agrees with the non-perturbative fixed point answer derived using anomaly arguments in~\cite{shi2022_gifts}. 

Having understood the boson self energy, we can then turn to the electrical conductivity. Recall from Section~\ref{subsec:surgery_suture_conductivity} that, using the braket notation, the current-current correlation function can be written as
\begin{equation}\label{eq:HM_1boson_GJJ}
    \begin{aligned}
    N^{-1} G_{J^i J^j}(\bs{q}=0,\Omega) &= \bra{v^i_{\bs{q}=0,\Omega}} W_G(1-K_{\rm MT} - K_{\rm AL})^{-1} \ket{v^j_{\bs{q}=0,\Omega}} \\
    &+ \bra{v^i_{\bs{q}=0,\Omega}} W_G (1-K_{\rm MT} - K_{\rm AL})^{-1} \ket{g_{\bs{q}=0,\Omega}} \\
    &\cdot D(\bs{q}=0,\Omega) \bra{g_{\bs{q}=0,\Omega}} W_G (1-K_{\rm MT} - K_{\rm AL})^{-1} \ket{v^j_{\bs{q}=0,\Omega}} \,. 
    \end{aligned}
\end{equation}
The tour de force calculation in~\cite{guo2023migdal,guo2023fluctuation} shows that the soft modes of $1-K_{\rm MT} - K_{\rm AL}$ correspond to near unit eigenvalues of $-W_G (1-K_{\rm MT} - K_{\rm AL})^{-1}$ (i.e. eigenvalues that approach 1 in the $\Omega \rightarrow 0$ limit). Moreover, these soft modes are generated by shape deformations of the Fermi surface. Assuming that the Fermi surface has only one connected component parametrized by an angle $\theta$, the space of all shape deformations is spanned by $\delta$-function bumps localized at every possible $\theta$. Within the spectral decomposition, the sum over these soft modes can be written as an integral over $\theta$ with an appropriate measure. From the boson self energy, we learned that the correct measure to choose for a generic Fermi surface shape is
\begin{equation}
    \sum_{\text{soft modes } \alpha} \ket{\alpha} \bra{\alpha} = \int d \theta \frac{\left|\partial_{\theta} \bs{k}_F(\theta)\right|}{(2\pi)^2 v_F(\theta)} \ket{\theta} \bra{\theta} \,.
\end{equation}
Summing up the soft mode contributions gives the most IR singular part of $G_{J_iJ_j}(\bs{q}=0, \Omega)$
\begin{equation}
    \begin{aligned}
        &- N^{-1} G_{J^i J^j}(\bs{q}=0,\Omega)\\
        &= \int d \theta \frac{\left|\partial_{\theta} \bs{k}_F(\theta)\right|}{(2\pi)^2\, v_F(\theta)}\, \bra{v^i_{\bs{q}=0,\Omega}}\ket{\theta} \bra{\theta} \ket{v^j_{\bs{q}=0,\Omega}} \\
        &+ \int d \theta \frac{\left|\partial_{\theta} \bs{k}_F(\theta)\right|}{(2\pi)^2 \,v_F(\theta)} \,\bra{v^i_{\bs{q}=0,\Omega}}\ket{\theta} \bra{\theta} \ket{g_{\bs{q}=0,\Omega}} \frac{1}{- \Pi(\bs{q} = 0, \Omega)} \int d \theta \frac{\left|\partial_{\theta} \bs{k}_F(\theta)\right|}{(2\pi)^2 \,v_F(\theta)}\, \bra{g_{\bs{q}=0,\Omega}}\ket{\theta} \bra{\theta} \ket{v^j_{\bs{q}=0,\Omega}} \\
        &= \Tr_{\theta} v^i\, v^j - \frac{\Tr_{\theta} (v^i \,g) \Tr_{\theta} (g\, v^j)}{\Tr g^2} + \delta G_{J^iJ^i}(\bs{q}=0, \Omega) \,, 
    \end{aligned}
\end{equation}
where $\delta G_{J^iJ^i}(\bs{q}=0, \Omega)$ is proportional to the frequency-dependent part of $\Pi(\bs{q} =0, \Omega)$. Plugging this correlation function into the conductivity then yields a formula for the total conductivity
\begin{equation}
    \sigma^{ij}(\bs{q}=0,\Omega) = \frac{i N}{\Omega} \left[\Tr_{\theta} v^i\, v^j - \frac{\Tr_{\theta} (v^i\, g) \Tr_{\theta} (g\, v^j)}{\Tr g^2} \right] + \delta \sigma^{ij}(\bs{q}=0, \Omega) \,,
\end{equation}
where
\begin{equation}
    \delta \sigma^{ij}(\bs{q}=0,\Omega) = \begin{cases}
        \mathcal{O}(\Omega^{(8-2z)/z}) & \text{convex FS} \\ \mathcal{O}(\Omega^{(4-2z)/z}) & \text{concave FS} 
    \end{cases} \,.
\end{equation}
Several comments are in order. First, the interaction-corrected Drude weight agrees precisely with non-perturbative arguments in~\cite{shi2022_gifts,shi2023_loop}. The form of the correction depends strongly on the interaction form factor $g(\theta) \equiv g(\bs{k}_F(\theta))$. In particular, since the Fermi velocity $v^i(\theta)$ contains only odd angular momentum modes, the Drude weight is only corrected when $g(\theta)$ also contains odd angular momentum modes. This is the case for loop current order but not for Ising-nematic order, as previously pointed out in~\cite{shi2023_loop}.

The structure of incoherent corrections is also interesting, as it shows a dichotomy between convex and concave Fermi surfaces. In this context, a one-dimensional curve (e.g. Fermi surface) is defined to be convex if for every point on the curve, there exists a line passing through the point such that the curve is contained in one of the two half-planes bounded by the line. Therefore, a circular Fermi surface is convex, while a starfish-shaped Fermi surface is concave. Why does concavity play a role in the scaling of optical transport? The rough intuition is that when a pair of virtual fermions are scattered by a virtual boson with momentum $\bs{q}$, the pair momenta change from $\bs{k} - \bs{q}/2, \bs{k}' + \bs{q}/2$ to $\bs{k} + \bs{q}/2, \bs{k}' - \bs{q}/2$. The energetically favorable scattering processes are those where all four fermions are nearly on shell (i.e.  $\epsilon_{\bs{k}-\bs{q}/2} \approx \epsilon_{\bs{k}+\bs{q}/2} \approx \epsilon_{\bs{k}'-\bs{q}/2} \approx \epsilon_{\bs{k}'+\bs{q}/2} = 0$). For a convex Fermi surface, there are at most two solutions to this equation related by inversion $\bs{k} \rightarrow -\bs{k}$. In these special inversion-related channels, the leading singular contributions from different diagrams cancel. However, for concave Fermi surfaces, there are more solutions that are not related by inversion. The contributions from these new channels generically do not cancel and preserve the singular scaling of individual diagrams. A more detailed discussion of this dichotomy between convex and concave Fermi surface can be found in a recent paper~\cite{Gindikin2024_convexconcave}.

So far, we have only accounted for the soft modes, which provide the leading Drude weight but not necessarily the most IR-singular incoherent corrections. Generically, when additional kinematic constraints emerge in the IR limit, the soft mode corrections in $\delta \sigma(\bs{q}=0, \Omega)$ can be anomalously suppressed and become less singular than the leading contributions from rough modes. This anomalous suppression does not happen for concave Fermi surfaces but does happen for convex Fermi surfaces. Since diagrammatic calculations is not the conceptual focus of our paper, we will not discuss these details any further, and refer the interested readers to a careful analysis of the competition between soft and rough modes in~\cite{guo2023migdal,guo2023fluctuation}.

\subsection{Generalizations to multiple boson flavors and gauge fields}\label{subsec:app_2boson}

The double expansion can also be generalized to Hertz-Millis models with multiple boson species and to certain beyond-Landau metallic quantum critical points with emergent gauge fields. The general idea is as follows: suppose we have a few species of fermions $f_i$ coupled to emergent gauge fields $A_i$ and critical bosons $\phi_i$. In a generalized double expansion, we introduce a non-analytic kinetic term $|\bs{q}|^{1+\epsilon}$ for all bosonic fields $A_i, \phi_i$ that couple to any of the Fermi surfaces $f_i$. Now for each fermion species, we introduce $N$ flavors labeled by $f_{i, \alpha}$ where $\alpha$ goes from $1 \sim N$. Then a controlled expansion can be developed again in the regime $\epsilon \rightarrow 0, N \rightarrow \infty$ with $\epsilon N$ fixed. 

With this generalized double expansion in mind, we can now ask if the mixed boson self energy matrix $\Pi_{\phi_i \phi_j}(\bs{q},\Omega)$ or mixed gauge field self energy matrix $\Pi_{A_i A_j}(\bs{q},\Omega)$ admits a tractable expansion in the transport regime $\Omega \gg |\bs{q}|$. Clearly, one can still define a notion of p-h irreducibility such that these self energy matrices can be recast as a geometric series. However, in the most general case where particles and holes come in multiple flavors and different bosons can mix up the flavors in different ways, the definition of p-h irreducibility is complicated and a useful classification scheme for p-h irreducible diagrams can only be developed on a case-by-case basis. Therefore, instead of attempting to make general statements, we will demonstrate how the classification works in a specific spinful Hertz-Millis model, where we can already appreciate both the complexities of additional flavors and the flexibility of the formalism developed in Section~\ref{sec:transport_expansion}. 

\begin{figure}[ht]
    \centering
    \includegraphics[width = \linewidth]{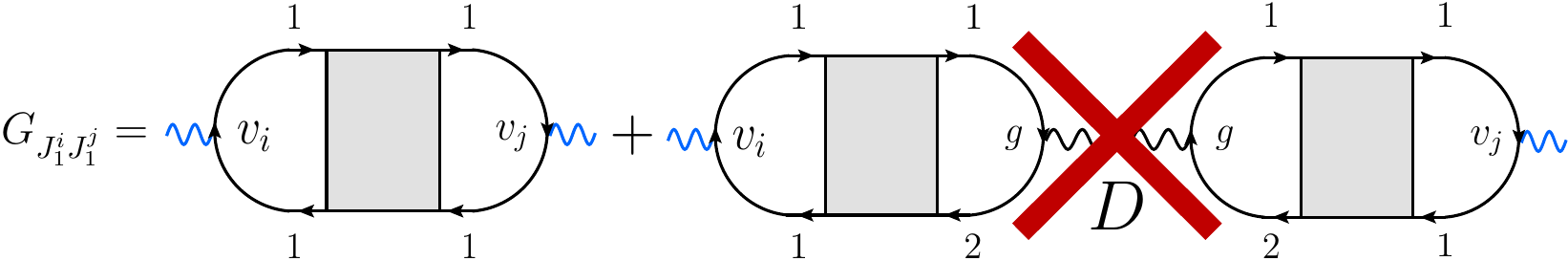}
    \caption{The naive decomposition of $G_{J_1^i J_1^j}$ into two terms misses the fact that the second term is identically zero.}
    \label{fig:GJJ_2boson}
\end{figure}

Consider a Fermi liquid formed by fermionic fields $f_{i, \alpha}$ where $i = 1, \ldots, N$ labels the different species and $\alpha = 1, 2$ labels the spin. The model we study describes a Landau ordering transition associated with the onset of XY ferromagnetic order in this metallic system. The XY order parameter is a two component bosonic field $(\phi_x, \phi_y)$ which couples to the fermion sector via an interaction term $\int \phi_x S^x + \phi_y S^y$ where the spin operators are defined by
\begin{equation}
    S^{x/y} = \frac{1}{2} \sum_{i=1}^N f^{\dagger}_{i,\alpha} \sigma^{x/y}_{\alpha\beta} f_{i,\beta} \,.
\end{equation}
Assuming that the ordering transition is continuous, the critical point can be captured by the following Euclidean action (within the generalized double expansion)
\begin{equation}\label{eq:HM_action_XY}
    \begin{aligned}
    S &= S_f + S_{\phi} + S_{\rm int} \,, \\
    S_f &= - \sum_{i = 1}^N \sum_{\alpha=1}^2 \int_{\bs{k}, \omega} f^{\dagger}_{i,\alpha}(\bs{k},\omega) \left[i\omega - \epsilon(\bs{k})\right] f_{i,\alpha}(\bs{k}, \omega)  \,, \\
    S_{\phi} &= \frac{1}{2} \int_{\bs{q}, \Omega} \sum_{a=x, y} \phi_a (\bs{q}, \Omega) \left[\lambda \Omega^2 + |\bs{q}|^{1+\epsilon}\right] \phi_a(\bs{q},\Omega)  \,, \\
    S_{\rm int} &= \frac{g}{2\sqrt{N}} \sum_{a=x, y} \sum_{\alpha, \beta = 1}^2 \sum_{i=1}^N \int_{\bs{r}, t} \, \phi_a(\bs{r}, t) \,f^{\dagger}_{i,\alpha}(\bs{r}, t) \, \sigma^a_{\alpha \beta} \,f_{i,\beta}(\bs{r}, t) \,.
    \end{aligned}
\end{equation}
We are interested in diagrammatic expansions for various bosonic correlation functions in the transport regime. The simplest such example is a spin-polarized current-current correlation function, which is related to the spin-polarized conductivity via the Kubo formula. Due to the exchange symmetry between spin labels $1 \leftrightarrow 2$, we can focus on the current-current correlator restricted to spin-1:
\begin{equation}
    \begin{aligned}
    G_{J_1^i J_1^j}(\bs{q} =0, \Omega) &= \ev{J_1^i(\bs{q}=0,\Omega) J_1^j(\bs{q}=0, -\Omega)} \,, \\
    J_1^i(\bs{q}) &= \sum_{m=1}^N \int \frac{d^2 \bs{k}}{(2\pi)^2} v_F^i(\bs{k}) \, f^{\dagger}_{m,1}(\bs{k} + \bs{q}/2) \, f_{m,1}(\bs{k}-\bs{q}/2) \,.
    \end{aligned}
\end{equation}
Following Section~\ref{subsec:surgery_suture_conductivity}, we might at first expect that $G_{J_1^i J_1^j}$ can be decomposed into two terms as in Fig.~\ref{fig:GJJ_2boson}.
The first term contains all diagrams that are irreducible with respect to internal boson line cuts and the second term contains the rest. However, the Yukawa interaction structure demands each internal vertex to be associated with an off-diagonal spin matrix $S^x$ or $S^y$. This means that the second term in Fig.~\ref{fig:GJJ_2boson} necessarily contains two subdiagrams that connect three fermion lines with spin index 1 and one fermion line with spin index 2. One can easily show that such diagrams vanish identically. Therefore, the second term does not contribute to the spin-polarized conductivity, as indicated by the red cross. 

Coming back to the first term, we just need to understand the grey box, which is a spin-polarized four-point fermion vertex. Slightly generalizing the arguments in Section~\ref{subsec:geometric_ph}, one can write this vertex as a geometric series of spin-1 particle-hole irreducible diagrams $K$.\footnote{Heuristically, this can be defined as all diagrams that cannot be factorized upon cutting a pair of spin-1 particle-hole lines. For a more precise definition, one can refer to Section~\ref{subsec:geometric_ph}.} 

The final step is to study the kernel $K$, which maps a pair of spin-1 particle-hole fermion lines to another pair. As far as $N$ counting is concerned, the surgery and suture procedures developed in Section~\ref{subsec:surgery_suture} still apply. However, the set of diagrams that contribute to $K$ at leading order in the double expansion will be different. This is because certain diagrams that have the correct $N$-scaling are forbidden by the interaction structure. For a simple example, consider the Maki-Thompson diagram which generates ladders in the spinless Hertz-Millis model. Since the Yukawa interaction is off-diagonal in spin space, the Maki-Thompson kernel maps a pair of spin-1 fermion lines to a pair of spin-2 fermion lines and vice versa. Hence the Maki-Thompson kernel cannot contribute to $K$. 

\begin{figure}[ht]
    \centering
    \includegraphics[width = \textwidth]{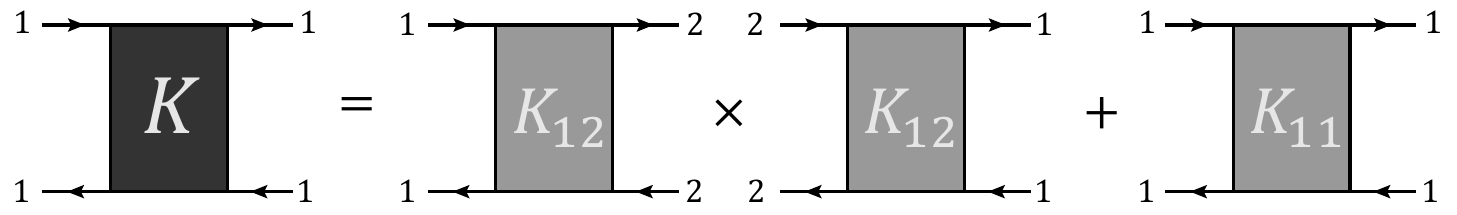}
    \includegraphics[width = 0.8\textwidth]{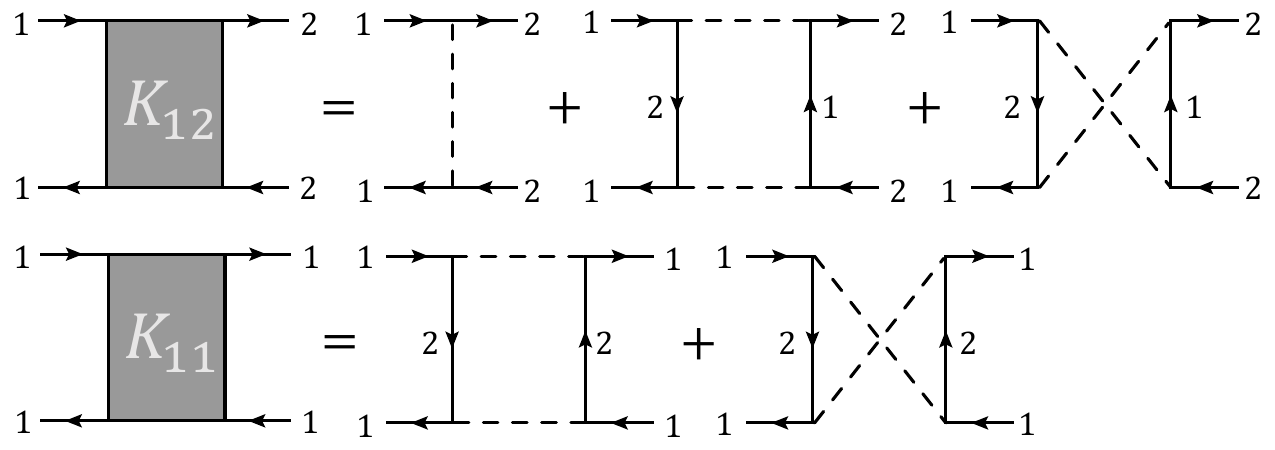}
    \caption{Diagrams that contribute to the kernel $K$ at leading order in the double expansion.}
    \label{fig:K_2boson}
\end{figure}

With these illegal diagrams removed, the remaining diagrams that contribute to the p-h irreducible kernel $K$ at $\mathcal{O}(N^0)$ can be written as $K_{11} + K_{12} (1 - K_{22})^{-1} K_{21}$ where $K_{ij}$ are kernels that maps between pairs of fermion lines with different/same spins. The full structure is shown in Fig.~\ref{fig:K_2boson}.
By the exchange symmetry between the two spin species, the spin labels attached to internal lines in the diagrams do not affect the functional forms of internal propagators. Therefore, we will write $K_{12} = K_{\rm MT} + K_{\rm AL}$ and $K_{11} = K_{22} = K_{\rm AL}$ and omit explicit spin labels. This decomposition immediately gives a compact formula for the spin-polarized current-current correlation function
\begin{equation}\label{eq:HM_2boson_GJJ}
    N^{-1} G_{J_1^i J_1^j}(\bs{q}, \Omega) = \bra{v^i_{\bs{q},\Omega}}W_G \left[1- (K_{\rm MT} + K_{\rm AL})(1 - K_{\rm AL})^{-1}(K_{\rm MT} + K_{\rm AL}) - K_{\rm AL}\right]^{-1} \ket{v^j_{\bs{q}, \Omega}} \,. 
\end{equation}
Let us compare this formula with the analogous formula \eqref{eq:HM_1boson_GJJ} that appears in the spinless Hertz-Millis model. The modification in the denominator has important physical consequences. From the analysis in the spinless case, we know that $1- (K_{\rm MT} + K_{\rm AL})(1 - K_{\rm AL})^{-1}(K_{\rm MT} + K_{\rm AL}) - K_{\rm AL}$ and $1 - (K_{\rm MT} + K_{\rm AL})$ have the same soft eigenvectors but different spectra. This shift in the soft mode spectra will not change the Drude part of the conductivity but can generically lead to a quantum critical incoherent conductivity in the spin-polarized channel. This is not surprising because the Yukawa interactions in this model violate spin conservation explicitly. Since this paper focuses on conceptual issues, we will not comment further on the detailed evaluation of \eqref{eq:HM_2boson_GJJ}, which is reported elsewhere~\cite{shi2024_exciton}.

This concludes our discussion of the spinful Hertz-Millis model which describes a metallic XY ferromagnetic transition. The most important takeaway is that the large $N$ counting structure developed in the spinless Hertz-Millis model survives in more general cases, although the definition of ``particle-hole irreducible kernel" and the set of diagrams that need to be included at each order in the double expansion can be modified in interesting ways. It is our hope that in every new problem of this type, one can use the general logic in Section~\ref{sec:transport_expansion} and~\ref{sec:application} to construct an appropriate classification of diagrams and perform systematic conductivity calculations. 

\section{Discussion}\label{sec:discussion}

Understanding the low energy properties of non-Fermi liquid metals is a longstanding challenge in condensed matter physics. The simplest class of non-Fermi liquids are described by Hertz-Millis models, where electronic modes near a Fermi surface are strongly coupled to the long wavelength fluctuations of gapless bosonic fields. In this work, we revisited a controlled ``double expansion"~\cite{Mross2010} for Hertz-Millis models, which combines an expansion in the inverse number $(N)$ of fermion flavors with an expansion in the boson dynamical critical exponent, $z = 2 + \epsilon$. While previous works studied correlation functions in the quantum critical regime where the external frequency $\Omega$ is much smaller than the external momentum $q$, we focused on the opposite regime $\Omega \gg q$ which is relevant for transport properties. The key challenge was that correlation functions receive contributions from a finite number of diagrams in the quantum critical regime, but an infinite number of diagrams in the transport regime \textit{at every order in the expansion}. Surprisingly, we show that this infinite set of diagrams can be reorganized into a geometric series $(1-K)^{-1}$ where the kernel $K$ only involves finitely many diagrams at each order in the expansion. Furthermore, we provide an explicit recipe for enumerating and resumming these diagrams that facilitates tractable calculations of the electrical conductivity. 

It is useful to zoom out and consider the benefits and drawbacks of the double expansion relative to other perturbative approaches in the literature.\footnote{Examples include the codimension expansion~\cite{DalidovichLee2013}, the matrix large $N$ expansion~\cite{Fitzpatrick2013,Raghu2015}, and the Yukawa-SYK expansion~\cite{Esterlis2021}. See~\cite{Lee2017_review} and Section 6 of~\cite{shi2022_gifts} for a brief review of the basic structure of each expansion scheme.} One important merit is that the double expansion preserves the microscopic $U(1)$ symmetry of the Hertz-Millis model and embeds it in a larger $U(N)$ symmetry of the deformed action. As a result, the metallic quantum critical point can be accessed by tuning a single $U(N)$-singlet relevant operator and the transport of conserved symmetry currents is well-defined at all stages of the calculation. This is to be contrasted with the codimension expansion~\cite{DalidovichLee2013} which explicitly breaks the microscopic $U(1)$ symmetry, and the Yukawa-SYK large $N$ expansion which is a multicritical point with a large number $(N^2)$ of relevant deformations (i.e. the boson mass matrix $R_{ij} \phi^i \phi^j$). A related advantage of the double expansion is that it preserves the anomaly structure of the physical theory with $N = 1, \epsilon = 1$. More precisely, at the IR fixed point, the electrical current associated with the $U(1)$ subgroup of $U(N)$ respects the same anomaly constraints for \textit{every value of $N$}. This feature is shared by the matrix large $N$ expansion but is violated in the Yukawa-SYK expansion as shown in~\cite{shi2023_loop}. Going beyond correlation functions that are constrained by symmetries and anomalies, the double expansion also provides an efficient calculational method that is competitive with the Yukawa-SYK expansion, as demonstrated for the incoherent optical conductivity in Section~\ref{subsec:app_1boson}. The analogous calculation has not been possible in the matrix large $N$ expansion, where a summation over planar diagrams may be required even at leading order.

The main caveat of the double expansion, as emphasized in~\cite{Ye2021}, is that certain four-loop diagrams for the boson self energy contain double-log divergences that cannot be cancelled by local counterterms. These divergences originate from virtual Cooper pairs spread across the entire Fermi surface, which are \textit{absent in all the other expansions where $\epsilon$ is fixed at 1}. However, it is currently unknown whether there exists other diagrams that can cancel these double-log divergences. This open question needs to be resolved before a complete assessment of the double expansion can be made. 

Looking ahead, it would be interesting to explore if the transport expansion developed in this paper can be applied to a wider class of non-Fermi liquids. One natural direction is the extension to more complicated Hertz-Millis models or beyond-Landau metallic quantum critical points that we briefly discussed in Section~\ref{subsec:app_2boson}. The presence of arbitrary interactions between multiple species of fermions and bosons generally weakens the constraints from non-perturbative anomalies and allows for critical fixed point incoherent conductivities that can potentially be accessed with a minor generalization of the double expansion. A more ambitious direction is to ask if the double expansion can be combined with some additional control parameter to study Hertz-Millis models with spatial disorder (see~\cite{Nosov2020_disorderNFL} for an interesting attempt using the matrix large $N$ expansion). Such a framework would clarify the interplay between interactions and disorder in the presence of a Fermi surface, which can lead to novel transport phenomena that are not possible with either of these ingredients alone. Finally, we must acknowledge that many non-Fermi liquids observed
in nature (including cuprates and heavy fermion metals) are likely not described by the Hertz-Millis model or any of its minor generalizations. Nevertheless, the essential lesson that the structure of diagrammatic expansions can be qualitatively different in distinct kinematic regimes is likely robust. We hope that this observation can serve as a useful guide in analyzing more realistic theories of non-Fermi liquids in the future. 



\section*{Acknowledgements}

The author thanks Luca Delacr\'etaz, Zhihuan Dong, Hart Goldman, Haoyu Guo, Sung-Sik Lee, Seth Musser, T. Senthil, Salvatore Pace, Weicheng Ye, and Liujun Zou for helpful discussions and feedback on the draft. This work is supported by the Department of Energy under grant DE-SC0008739.

\appendix

\section{Proofs of two claims in Section~\ref{subsubsec:surgery_suture_explained}}\label{app:proofs}

In this appendix, we prove two claims that are crucial for the classification of particle-hole irreducible diagrams. Let $K_i$ be a diagram that appears in the expansion of the p-h irreducible kernel $K$ and let $\Pi_{\rm amp,i}$ be the diagram that remains after we apply the ``surgery + reduction" procedure to $K_i$ as defined in Section~\ref{subsec:surgery_suture}. Then the following claims are true: \newline

\noindent \textbf{Claim 2}: $\Pi_{\rm amp,i}$ is not boson-$n$-reducible for any $n > 1$.
\begin{proof}
    Recall from Section~\ref{subsec:surgery_suture} that a diagram is \textbf{boson-$n$-reducible} if one can factorize the diagram into $n+1$ disconnected subdiagrams by cutting $n$ internal boson lines. 
    We prove that $\Pi_{\rm amp,i}$ cannot be boson-2-reducible. This would imply that $K_i$ is not boson-$n$-reducible for any $n \geq 2$ because each additional cut in a graph can only generate one additional disconnected component.

    Suppose for the sake of contradiction that $\Pi_{\rm amp,i}$ is boson-2-reducible. Then the original diagram $K_i$ can be factorized into 3 disconnected components by three boson line cuts. Now we do a casework.

    Case 1: If $K_i$ is a horizontal p-h irreducible diagram, then $K_i$ cannot contain any bSE subdiagram by the assumption that $K_i$ is a skeleton diagram. Therefore, it is impossible to factorize $K_i$ with 2 cuts. But in any Feynman graph, each additional cut can generate at most one extra disconnected component. Hence, 3 cuts can at most factorize the diagram into 2 disconnected components, leading to a contradiction. 

    Case 2: If $K_i$ is a vertical irreducible diagram, the situation is a little trickier because although the open four-fermion vertex does not contain any bSE subdiagram, one bSE subdiagram could appear when we close the four external fermion legs in pairs, generating a new diagram $\tilde K_i$ with two extra fermion loops (this possibility illustrated for a simple example in Fig.~\ref{fig:Claim2_diagram}).

    \begin{figure}[ht]
        \centering
        \includegraphics[width = 0.8\linewidth]{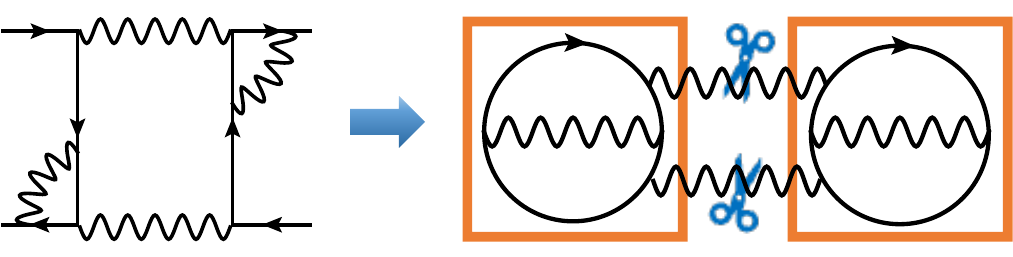}
        \caption{Take the vertical p-h irreducible kernel $K_i$ on the left. When the four external fermion legs are sewed together into loops, we get a new diagram which does contain a bSE subdiagram. In this case, two cuts indeed factorize $K_i$ into two disconnected components as indicated by the orange boxes.}
        \label{fig:Claim2_diagram}
    \end{figure}
    
    If we are only allowed to cut internal boson lines that are not connected to the external fermion bubbles, then the same argument as in case 1 gives a contradiction. If instead we cut internal boson lines that are connected to the external fermion bubbles, then in order to factorize $\tilde K_i$ into two pieces with two cuts, these two cuts must be the boson lines in the immediate vicinity of one of the external fermion bubbles. Can the remaining diagram be factorized with a single additional boson line cut? The answer is negative, because the existence of such a cut contradicts the condition that the open kernel $K_i$ is irreducible with respect to a single boson line cut. Thus, three cuts cannot give three disconnected components, leading again to a contradiction. 
\end{proof}

\noindent \textbf{Claim 3}: $\Pi_{\rm amp,i}$ does not contain a fermion self energy (fSE) subdiagram. 
\begin{proof}
    We say that a diagram contains an fSE subdiagram if we can extract a disconnected piece from the diagram by cutting two internal fermion lines. Now imagine that we cut a boson line from $K_i$ and denote the exposed boson legs by $a$ and $b$. Suppose for the sake of contradiction that $\Pi_{\rm amp,i}$ contains a fermion loop $L$ which hosts an fSE subdiagram $\Sigma$ (this basic configuration is shown in Fig.~\ref{fig:Claim3_diagram}).
    \begin{figure}[ht]
        \centering
        \includegraphics[width=\linewidth]{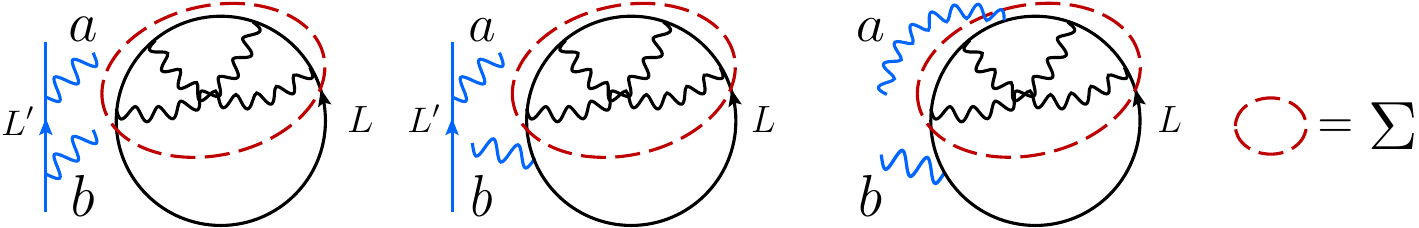}
        \caption{Three possible scenarios for the intertwinement of the fermion self energy subdiagram and the exposed boson legs $a,b$. On the left, $a,b$ are disjoint from the loop $L$. In the middle, $a$ is attached to loop $L'$ while $b$ is attached to loop $L$. On the right, $a,b$ are both attached to loop $L$.}
        \label{fig:Claim3_diagram}
    \end{figure}

    If the exposed boson legs are completely disconnected to the fermion loop $L$ or if one of the dangling boson legs is connected to $L$ while the other dangling boson leg is connected to some other fermion loop $L'$, then upon reconnecting $a$ and $b$, the original diagram $K_i$ would also contain $\Sigma$, leading to a contradiction (see left/middle panel of Fig.~\ref{fig:Claim3_diagram}). If both of the exposed boson legs connect to the fermion loop $L$, then the only potential concern is if one of the legs penetrates into the fSE subdiagram (see right panel of Fig.~\ref{fig:Claim3_diagram}). But such a configuration contradicts the assumption that $\Sigma$ is an fSE subdiagram of $\Pi_{\rm amp, i}$. 
    
    The above argument demonstrates that $K_i$ contains an fSE subdiagram whenever $\Pi_{\rm amp, i}$ does, which contradicts the fact that $K_i$ is a skeleton diagram. 
\end{proof}

\nocite{apsrev41Control}
\bibliographystyle{apsrev4-1}
\bibliography{Mross_transport}

\end{document}